\documentclass[conference]{IEEEtran}
\IEEEoverridecommandlockouts

\usepackage{algorithm, algpseudocode}
\usepackage{cite}
\usepackage{amsmath,amssymb,amsfonts}
\usepackage{graphicx}
\usepackage{textcomp}
\usepackage[pdftex]{xcolor}

\usepackage{tabularx} 
\usepackage[frozencache=true,cachedir=minted-cache]{minted}
\usepackage{comment}
\usepackage[disable]{todonotes} 

\def\BibTeX{{\rm B\kern-.05em{\sc i\kern-.025em b}\kern-.08em
    T\kern-.1667em\lower.7ex\hbox{E}\kern-.125emX}}

\newcommand\angcmtfix[1]{\todo[inline, size=\small, color=green!10]{[fixed] Ang: #1}}

\usepackage{graphics}
\usepackage{subcaption}
\usepackage[normalem]{ulem}
\useunder{\uline}{\ul}{}
\usepackage{array}

\usepackage[utf8]{inputenc}
\usepackage[english]{babel}
\usepackage[]{caption}
\usepackage{multirow}

\definecolor{LightGray}{gray}{0.9}

\begin{document}

\title{Bit-GraphBLAS: Bit-Level Optimizations of Matrix-Centric Graph Processing on GPU}

\author{
\IEEEauthorblockN{
Jou-An Chen\IEEEauthorrefmark{1}, Hsin-Hsuan Sung\IEEEauthorrefmark{1}, Xipeng Shen\IEEEauthorrefmark{1}, Nathan Tallent\IEEEauthorrefmark{2}, Kevin Barker\IEEEauthorrefmark{2}, and Ang Li\IEEEauthorrefmark{2}
}
\IEEEauthorblockA{
Department of Computer Science, North Carolina State University, Raleigh, NC, USA\IEEEauthorrefmark{1}\\
Pacific Northwest National Laboratory, Richland, WA, USA\IEEEauthorrefmark{2}\\
\IEEEauthorrefmark{1}\{jchen73, hsung2, xshen5\}@ncsu.edu, \IEEEauthorrefmark{2}\{nathan.tallent, kevin.barker, ang.li\}@pnnl.gov
}
}

\maketitle
\pagestyle{plain}

\begin{abstract}
In a general graph data structure like an adjacency matrix, when edges are homogeneous, the connectivity of two nodes can be sufficiently represented using a single bit. This insight has, however, not yet been adequately exploited by the existing matrix-centric graph processing frameworks. This work fills the void by systematically exploring the bit-level representation of graphs and the corresponding optimizations to the graph operations. It proposes a two-level representation named Bit-Block Compressed Sparse Row (B2SR) and presents a series of optimizations to the graph operations on B2SR by leveraging the intrinsics of modern GPUs. Evaluations on NVIDIA Pascal and Volta GPUs show that the optimizations bring up to $40\times$ and $6555\times$ for essential GraphBLAS kernels SpMV and SpGEMM, respectively, making GraphBLAS-based BFS accelerate up to $433\times$, SSSP, PR, and CC up to $35\times$, and TC up to $52\times$. 
\end{abstract}


\section{Introduction}


Despite drawing great attention recently \cite{khorasani2014cusha, wang2016gunrock, zhang2016gbtl, yang2020graphblast, beamer2012direction, liu2015enterprise, merrill2012scalable, davidson2014work, wu2010efficient, soman2010fast, bisson2017high}, accelerating graph analytics on GPUs remains challenging in that: (i) graphs are often irregular in connectivity, causing warp divergence, memory non-coalescing, and low arithmetic intensity;  (ii) graphs are often large, many of which cannot directly fit into the caches, shared memory, or even the DRAM of GPUs. To address these challenges, the matrix-centric approach has been proposed and increasingly employed by GPU-based graph frameworks~\cite{nvgraph2020, SuiteSparse, zhang2016gbtl, yang2020graphblast}. Unlike traditional graph-centric frameworks that go through relevant nodes or edges iteratively, this approach employs sparse storage formats---such as compressed sparse column (CSC) or compressed sparse row (CSR)---to represent the adjacency matrix of a graph and then uses highly optimized linear algebra kernels (sparse matrix-vector and matrix-matrix multiplications, SpMV and SpGEMM) on the sparse formats for computation. 

Although GraphBLAS has shown improved performance over traditional approaches in handling large graphs, there is still considerable potential to tap into. This work aims to unlock the potential by exploiting bit-level representations and optimizations on GPUs. The underlying observation is that for a large class of graphs (homogeneous graphs), a single bit is sufficient for indicating the adjacency relation of two vertices in a graph (1: adjacent, 0: not adjacent). Given that many graph algorithms center around computations upon adjacency matrices, using a bit-level representation can potentially reduce storage usage and improve computation efficiency.

Bit representations (bitmaps, bitvectors) have been used in vertex-based graph frameworks~\cite{sundaram2015graphmat, GraphIt,orachev2021spbla} for representing frontiers (i.e., active nodes); bit-level optimizations have, however, not yet been systematically explored in matrix-based graph frameworks. Existing GraphBLAS~\cite{kepner2016mathematical, mattson2017graphblas} frameworks typically build on existing linear algebra libraries, which offer no bit-level representations of matrices or bit-level implementations of linear algebra functions. 



This work answers three key research questions. 

{\em \textbf{RQ-1}: What storage format should be used for a binary adjacency matrix?}

Unlike the Boolean data structures (frontiers) in graph-centric frameworks, for matrix-based graph frameworks, the binary data structure in focus is the entire adjacency matrix, which sits at the center of the linear algebra operations in GraphBLAS. Systematic studies are needed for manipulating it; simply representing it as a bitmap cannot tap into the full potential of space savings by accommodating many unnecessary zeros; that also causes difficulties for the graph operations to leverage the highly tuned matrix-based libraries.

Based on the properties of adjacency matrices and various tradeoffs, we design a storage format, namely Bit-Block Compressed Sparse Row (B2SR). B2SR is inspired by the Block Compressed Sparse Row (BSR) format~\cite{im1999optimizing}. It takes a two-level structure: The upper level is similar to BSR's upper level, using a sparse format to represent the locations of non-zero blocks (or called submatrices);  the lower level differs from BSR in that it represents each non-zero block as a dense {\em bit} matrix---that is, each element in the block becomes one bit in the representation. The representation allows it to efficiently harvest the hardware computation capability at the low level and at the same time leverage the (regional) sparsity at the high level. Balancing the space savings and the indexing overhead lays the foundation for tapping into the potential of sparse binary matrices.

{\em \textbf{RQ-2}: How to efficiently compute on the new representation?}

We propose several new algorithms to implement the critical linear algebra kernels (SpMV and SpGEMM) for manipulating sparse matrices represented in B2SR. The design carefully tailors the kernel implementations around the efficient low-level bit-manipulation intrinsics on GPUs. It exploits the new optimizations and hardware-specific capability brought by the latest GPUs. These kernels lay the foundation for efficient manipulations of sparse binary matrices for graph analytics.
\\

{\em \textbf{RQ-3}: What are the performance implications?}

We evaluate the B2SR-based SpMV and SpGEMM on 521 binary matrices and five graph algorithms. The benefits are significant. B2SR provides up to $32\times$ space savings. On two generations of GPUs (NVIDIA's Pascal and Volta), we observe 40$\times$ and 6555$\times$ maximum speedups over the state-of-the-art sparse linear algebra libraries cuSPARSE~\cite{cusparse2020} and GraphBLAST~\cite{yang2020graphblast}. On graph algorithms, it offers up to $433\times$ acceleration on Breadth-first-search (BFS), $55\times$ on Single-Source-Shortest-Path (SSSP), $28\times$ on PageRank (PR), $69\times$ on Connected Component (CC) algorithms, and $52\times$ on Triangle Counting (TC) algorithm over GraphBLAST~\cite{yang2020graphblast}, a GPU graph processing framework with state-of-the-art performance. As is well known ~\cite{zhao2018overhead, zhao2018bridging, zhou2019enabling}, no sparse format fits all matrices. So despite the effectiveness of B2SR, there are matrices that fit other sparse formats better. We provide a brief discussion and a simple sampling approach to assisting users in applying B2SR.

\section{Background and Related Work}


Graph programming frameworks are based on either graph-centric abstraction (vertex- or edge-centric)~\cite{fu2014mapgraph, khorasani2014cusha, malewicz2010pregel, mccune2015thinking, nguyen2013lightweight, roy2013x} or matrix abstraction. For the performance advantages and direct leverage of advances in high-performance linear algebra libraries, matrix abstraction has received increasing interest in recent years. GraphBLAS~\cite{kepner2016mathematical, mattson2017graphblas} is the mathematical core of matrix-based graph frameworks. It models graph traversal as operations on semi-rings. Frameworks that implement the standard include nvGraph~\cite{nvgraph2020}, cuGraph~\cite{cugraph}, SuiteSparse~\cite{SuiteSparse}, GraphBLAS template library (GBTL)~\cite{zhang2016gbtl}, GraphBLAST~\cite{yang2020graphblast}, and so on. Among them, GraphBLAST~\cite{yang2020graphblast} represents state of the art, achieving high performance on GPU by exploiting input and output sparsity~\cite{yang2018implementing} and enhanced load balance by exploiting the memory access patterns of sparse matrix multiplication.

These frameworks are mainly built under the same line of unified graph construct--using CSC or CSR to establish floating-point element space and perform matrix operations with underlying linear algebra libraries. They have not exploited bit-level optimizations. Even though in the sparse linear algebra libraries, bitmaps or bitvectors may be used to index the non-zero elements in a sparse matrix~\cite{buluc2011reduced,tang2013accelerating,zachariadis2020accelerating}, those libraries fundamentally assume that the sparse matrices are general rather than binary matrices. They hence leave an immense performance potential untapped (as our comparison in Section~\ref{sec:eval} shows). 


In graph-centric frameworks (e.g., GraphMat~\cite{sundaram2015graphmat}, Graphlt~\cite{GraphIt}, SpbLA~\cite{orachev2021spbla}), there are some Boolean data structures (e.g., frontiers or active nodes), which are sometimes represented in bitmaps or bitvectors. Some works~\cite{li2016compression,besta2018log,brisaboa2009k} exploit binary encoding or compression to achieve storage reduction and algorithm acceleration. For matrix-centric graph frameworks, the binary data structure in focus is the entire adjacency matrix, which sits at the center of the linear algebra operations in GraphBLAS---simply representing it as a bitmap leaves not only lots of potential for space savings but also causes difficulties for the graph operations to leverage the highly tuned matrix-based libraries. Therefore, systematic studies are needed for manipulating it. 

Bit-manipulation primitives have been used in enhancing the performance of deep neural networks (DNNs) on GPUs~\cite{BSTC, li2020accelerating, feng2021apnn}, ASICs, and FPGAs~\cite{ghodrati2020bit, sharma2018bit, eckert2018neural}. They are about dense binary operations, so they cannot efficiently handle sparse ops as those in graph processing. The irregularity of sparse matrices and accesses makes the issue more complex.

\section{Representation: B2SR}
\begin{figure}
    \centering
    \includegraphics[trim={0cm 20cm 15cm 0cm},clip,width=0.55\textwidth]{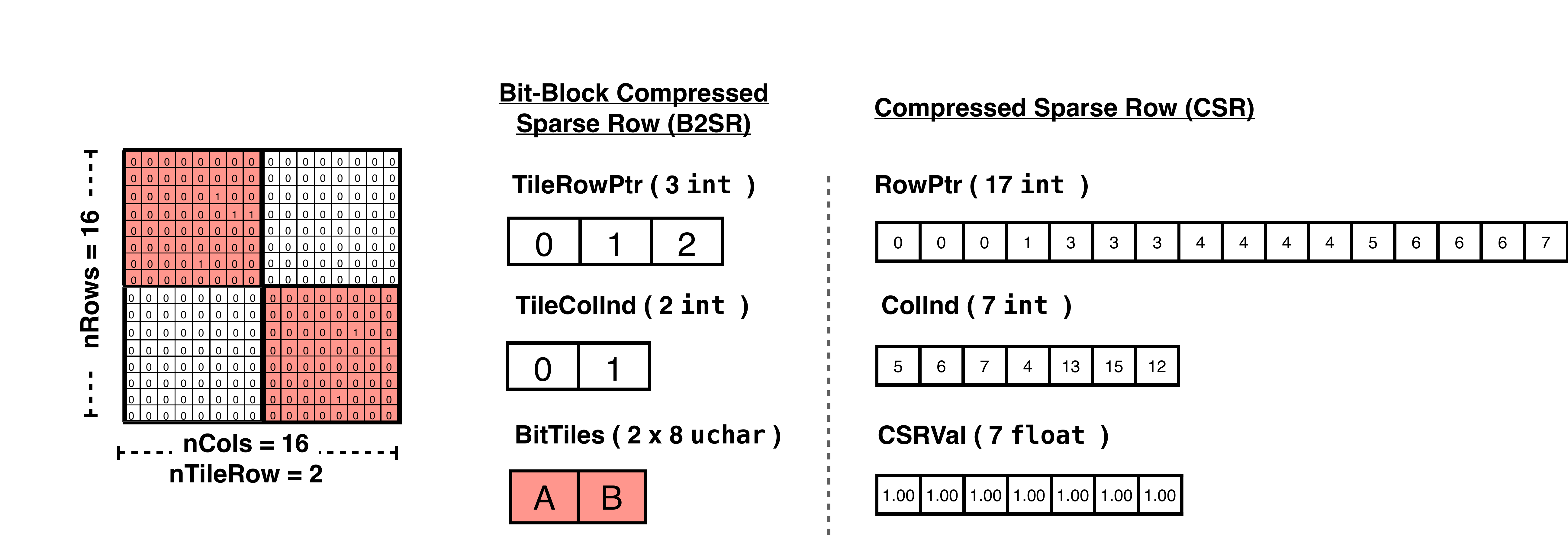}
    \caption{Illustration of the Bit-Block Compressed Sparse Row (B2SR).}
    \label{fig-format}
\end{figure}


This section presents B2SR, the format we have designed for representing an adjacency matrix. Figure~\ref{fig-format} illustrates the design. The principle we followed is that the representation should reduce the space cost as much as possible and at the same time facilitate the acceleration of the core graph operations. Drawing on the inspiration of BSR, we create the two-level representation of B2SR. The top-level takes advantage of well-proven effective sparse formats (CSR or CSC) to represent non-empty blocks. The bottom-level treats each non-empty block as a dense block and packs its elements into a bit representation. The combination of sparse formats and bit representation minimizes space usage, while the block-level dense bit format preserves low-level regularity, making efficient computation possible. We will explain the format in detail.


\subsection {Bit-tile Indexing System}
Since adjacency matrices are all square, it is natural to have the number of tile rows (\textit{nTileRow}) set as $\frac{nRows + tileDim - 1}{tileDim}$, where $tileDim$ is the dimension (or bit-width) of the tile (e.g., 4, 8, 16, 32). The number of non-empty bit-tiles can then be inferred from the nonzeros' coordinates of the sparse matrices.  In our implementation, we use cuSPARSE's \textbf{\texttt{cusparseXcsr2bsrNnz()}} API to obtain the number of non-empty tiles from the CSR format. We utilize indexing arrays to record the coordination of the non-empty byte-aligned bit-tiles. The proposed format comprises three arrays: 

\begin{itemize}
\item Tile row indices (\textit{TileRowPtr}): an integer array with the size of the number of tile-rows. It records the bit-tiles' row indices. It is an accumulated array with the i-th element recording the sum of total non-empty bit-tiles counting from the first tile row to the (i-1)-th row. Therefore, $TileRowPtr[i+1] - TileRowPtr[i]$ suggests the number of non-empty bit-tiles in the i-th tile-row.
\item Tile column indices (\textit{TileColInd}): an integer array with the size equal to the number of non-empty bit-tiles. This array is for recording the tile column indices in the tile coordination system.
\item Bit-tiles storage (\textit{BitTiles}): a bit-packing type (unsigned char, unsigned short, unsigned int, or unsigned long long int) array with a size equal to the \textit{tileDim}$\times$\textit{numofTiles} (number of non-empty bit-tiles). It stores the binarized non-empty bit-tiles' layout in the order of their tile column indices.
\end{itemize}

The proposed format has several merits: (1) It allows simpler transpose of the sparse matrix. By transforming the \textit{TileRowPtr} and \textit{TileColInd} from CSR to CSC, the sparse matrix is transposed. (\textit{BitTiles} do not require transpose since we default the \textbf{\texttt{mxv()}} and \textbf{\texttt{mxm()}} algorithm to access the content of a tile always in row-by-row order.) We use cuSPARSE's \textbf{\texttt{cusparseScsr2csc()}} API to enable this function in our implementation. (2) The format has a better data accessing locality for SpGEMM and SpMV when computing in tile-row by tile-row order; Since the storage format is similar to CSR, we can use existing optimization on CSR-based algorithms to spearhead the linear algebra kernels in use. (3) In \textit{BitTiles}, the format provides storage-saving compared to CSR. The proposed format carries extra zeros than CSR and COO, which store only the nonzeros. Nevertheless, with the binarized packing yielding up to $32\times$ space-saving per tile, most sparse matrices can still benefit from storage compression, especially when configured in small tile size.

\subsection {Bit Packing}
\begin{figure}
    \centering
    \includegraphics[width=0.5\textwidth]{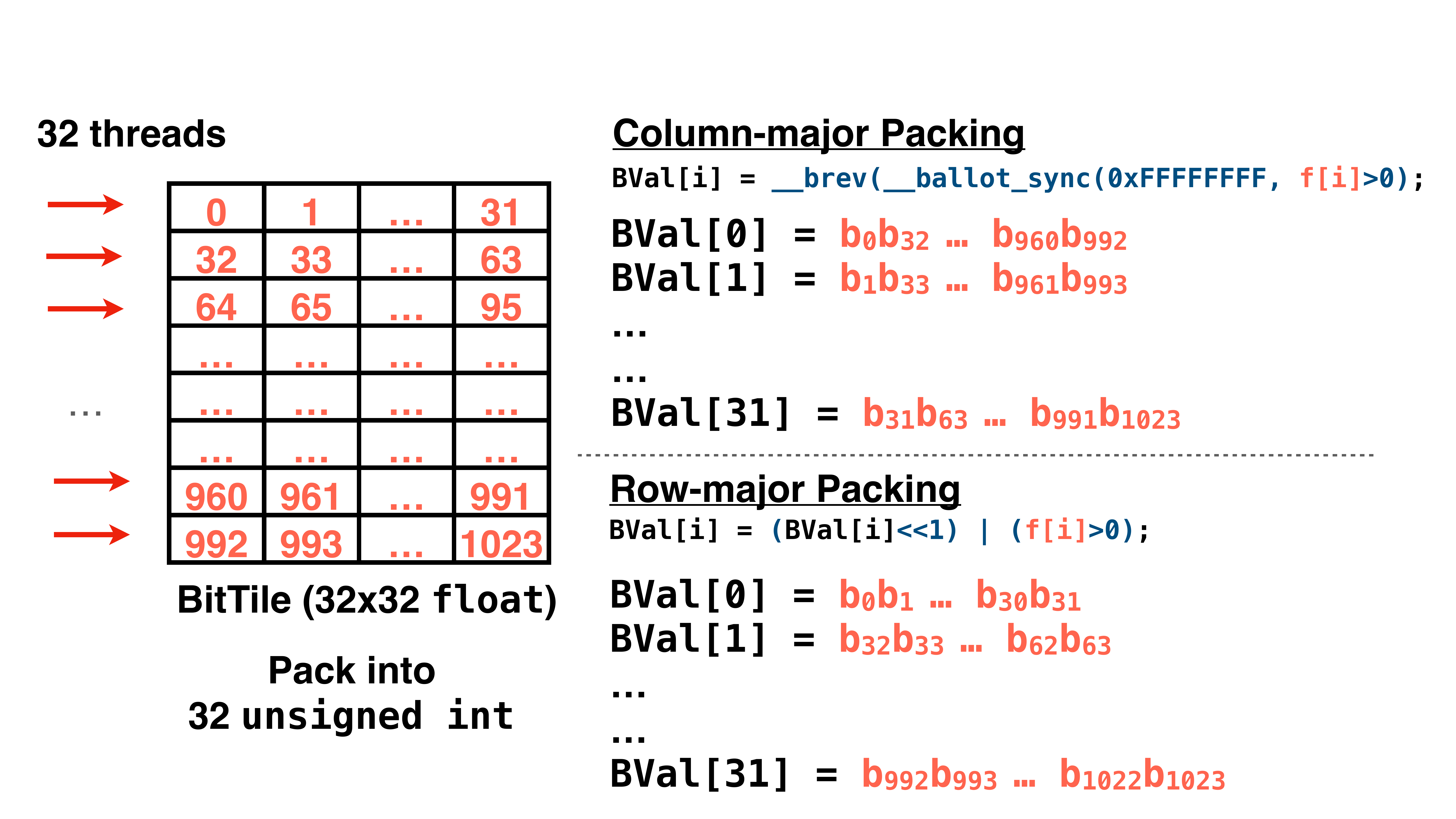}
    \caption{Illustration of column-major and row-major bit packing.}
    \label{fig-pack}
\end{figure}

\begin{table}[]
\caption{Binarized packing format.}
\label{tab-packing-format}
\centering
\resizebox{\columnwidth}{!}{
\begin{tabular}{|c|c|c|c|}
\hline
\textbf{Tile Size} & \textbf{\begin{tabular}[c]{@{}c@{}}CSR Storage\\ (at most)\end{tabular}} & \textbf{Binarized Packing}          & \textbf{\begin{tabular}[c]{@{}c@{}}Space Saving\\ per Tile\end{tabular}} \\ \hline
$4\times4$         & $4\times4$ \textbf{\texttt{float}}                                                & $4\times1$ \textbf{\texttt{unsigned char}}   & ~$16\times$                                                              \\ \hline
$8\times8$         & $8\times8$ \textbf{\texttt{float}}                                                & $8\times1$ \textbf{\texttt{unsigned char}}   & ~$32\times$                                                              \\ \hline
$16\times16$       & $16\times16$ \textbf{\texttt{float}}                                              & $16\times1$ \textbf{\texttt{unsigned short}} & ~$32\times$                                                              \\ \hline
$32\times32$       & $32\times32$ \textbf{\texttt{float}}                                              & $32\times32$ \textbf{\texttt{unsigned int}}  & ~$32\times$                                                              \\ \hline
\end{tabular}
}
\end{table}

Figure~\ref{fig-pack} presents the column-major and row-major packing of a bit-tile. We adopt \textit{column-major packing} as default when converting CSR to B2SR. The transpose of B2SR tiles can be achieved by storing the additional row-major layouts. To figure out the proper byte-addressable data types to carry bits in proximity, we explore the packing granularity from 8-bit to 32-bit (reference Table~\ref{tab-packing-format}). Additionally, we use half of the space in an \textbf{\texttt{unsigned char}} to allow 4-bit (nibble) packing, which further reduces half in one dimension to carry unnecessary zero when the matrix is extremely sparse. The packing result yields four variants of B2SR format: B2SR-4 (for 4$\times$4 tile size), B2SR-8 (for 8$\times$8 tile size), B2SR-16 (for 16$\times$16 tile size), and B2SR-32 (for 32$\times$32 tile size). Generally, the space-saving depends on how the adjacency matrix is initially being stored. State-of-the-art GPU graph frameworks mostly use \textbf{\texttt{float}} to carry the elements, so our bit-packing can generally provide up to 32$\times$ storage savings per square tile; for frameworks that use \textbf{\texttt{double}} to carry the elements, the bit-packing results in up to 64$\times$ savings in space. This indicates an ability to store 32$\times$ or 64$\times$ larger graphs using the same amount of space. This also brings a potential reduction in required data accessing bandwidth during computation to enable higher throughput.

\textbf{Bit-packing overhead}
To transform CSR to B2SR, we parallelize each tile-row's encoding procedure for the large graph. For the 4, 8, 16, or 32 continuous elements in CSR's \textbf{\texttt{RowPtr}}, we use \textbf{\texttt{cusparseXcsr2bsrNnz()}} and \textbf{\texttt{cusparseScsr2bsr()}} to obtain tile-row index and full-precision tiles along a tile-row. Next, column-major or row-major bit-packing kernels are applied to process each tile's encoding. The routine's overall cost is about 3 to 34 ms. In practical applications, a graph can be reused by many users; even within one execution, a graph is often used repeatedly (e.g., for iterative processing). So despite format conversion may be needed, such a one-time cost can be greatly amortized in these conditions.

\subsection{Sampling Profile and Tile Size Configuration}
\begin{figure}
    \begin{subfigure}[b]{.24\textwidth}
    \centering
    \includegraphics[scale=.24,width=\textwidth]{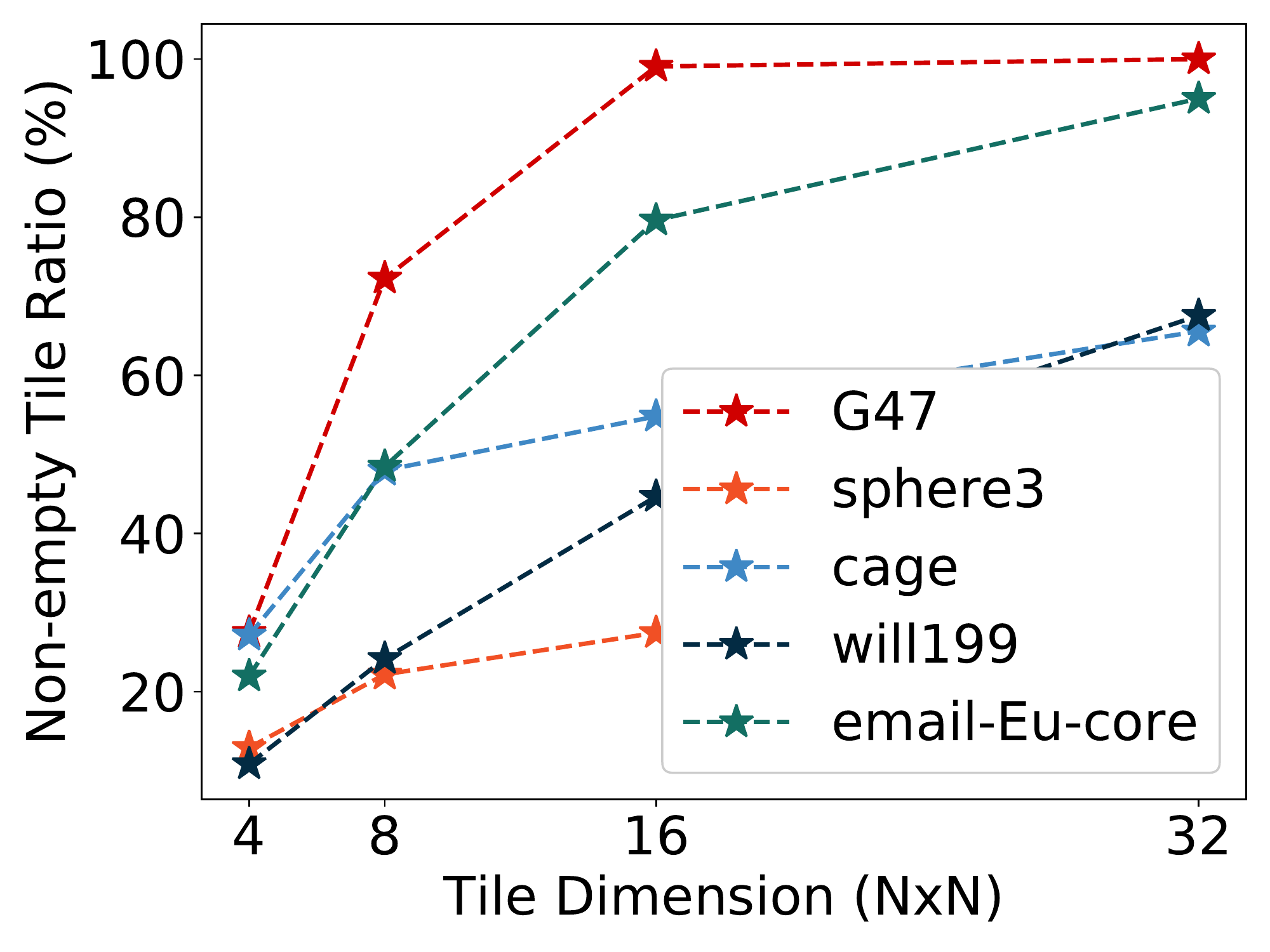}
    \caption{Tile proportion trend.}\label{fig-tile-percent}
    \end{subfigure}
    \begin{subfigure}[b]{.24\textwidth}
    \centering
    \includegraphics[scale=.24,width=\textwidth]{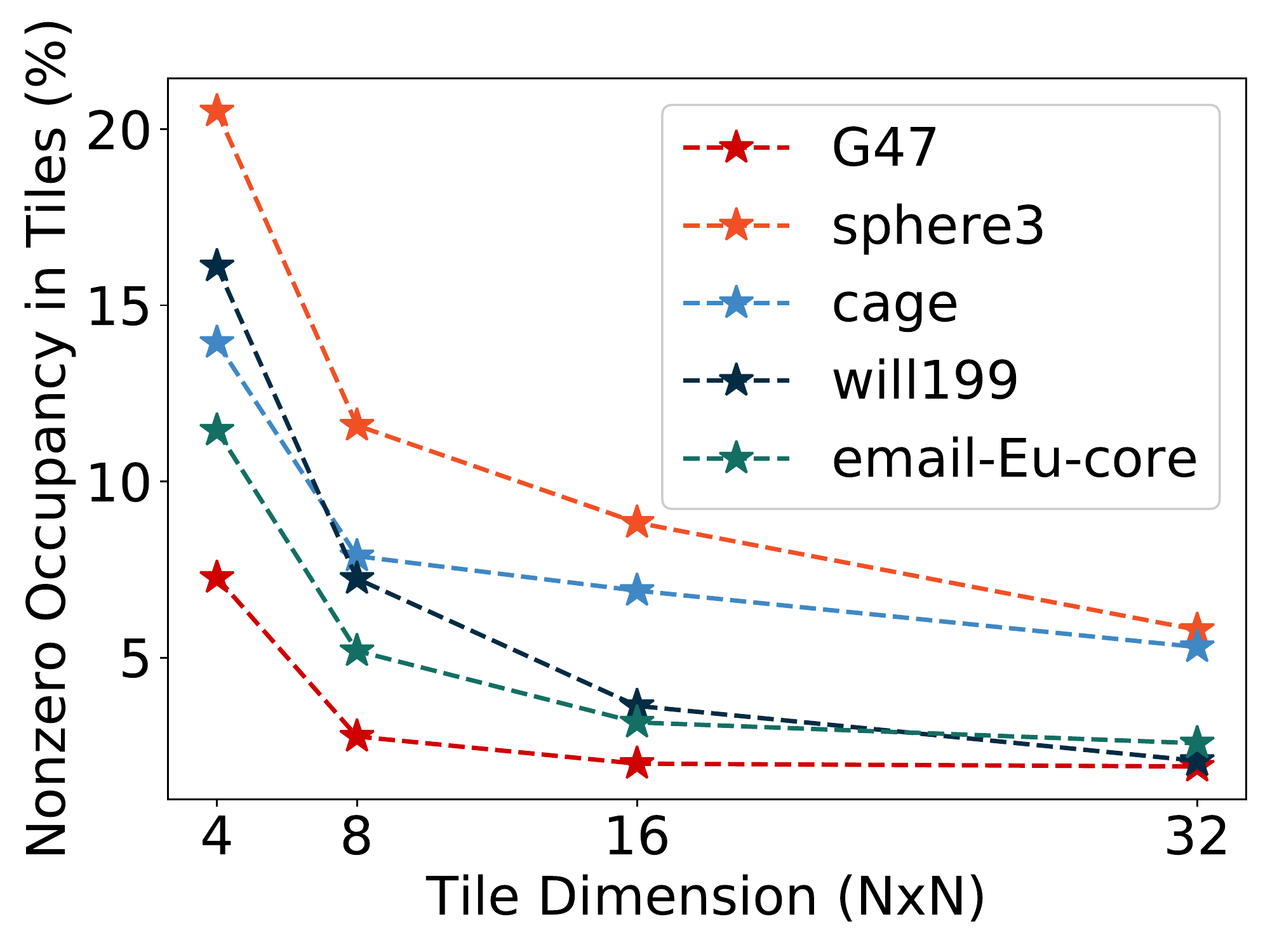}
    \caption{Nonzero occupancy trend.}\label{fig-tile-nnz}
    \end{subfigure}
    \caption{Effect trends with increment in tile size dimension.}\label{fig-tile}
\end{figure}

While the proposed method provides storage compression and performance gains in BLAS operators, it is evident that not all graph matrices are suitable for converting to Bit-GraphBLAS binarized format. For example, graph matrices with randomly distributed connections (nonzeros) or relatively dense patterns may not necessarily benefit from this format. When each bit tile does not capture enough nonzeros, we can have many empty bit-rows in tiles after bit packing. Converting the matrices from CSR to B2SR is not ideal since it expands the total amounts of storage. In addition, it adds additional per-thread workloads (OPS) compared to CSR's BLAS kernels in terms of operators.
We observe that the tile size selection trade-off can be considerably different in various matrix patterns. Figure~\ref{fig-tile-percent} shows that when the tile size equals 4$\times$4, there are less than 30$\%$ non-empty tiles; when the tile size equals 32$\times$32, non-empty tiles reach more than 80$\%$ for some matrices. The reason is that although increasing tile size can decrease the number of non-empty tiles, the reduction is often less than 4$\times$ (the times of per tile size increment), causing the ratio of non-empty tiles to increase ultimately. We find that despite the increment, the total B2SR byte size can sometimes decrease because of the reduction in the number of tiles and indexing arrays. For instance, in matrix \textit{mycielskian12}, we have CSR storage at 3.12 (MB), B2SR-4 at 675.70 (KB), B2SR-8 at 361.46 (KB), B2SR-16 at 358.89 (KB), and B2SR-32 at 429.89 (KB). The total byte size of the format does not monotonically increase as tile size increases. From the other angle, Figure~\ref{fig-tile-nnz} presents the reduction in the average occupancy of nonzeros in the non-empty tiles. The percentage of actual nonzeros in tiles can drop from 20$\%$ to less than 5$\%$ as the tile dimension differs. If the bit-tile is too large, the computation may waste processing too many tiles; if the bit-tile is too small, the indexing array may carry more unit workloads. 
Therefore, to fully utilize the double benefits (compression and computation efficiency) of Bit-GraphBLAS, graph users can first identify the potential benefits offline through our sampling profile method. In this way, when tackling extensive graphs or cumbersome, repetitive graph computations, users can experience worthy time and labor cost saving with only one-time format conversion to B2SR.
The sampling works as in Algorithm~\ref{sampleprofile}. The user first specifies the number of rows to sample as $N$ for $N \leq NumOfRows$ (sampling more rows can accurately capture matrix characteristics but induces more significant overheads). We then have $N$ random indices as the random index set $S \subseteq [0, 1, 2, ..., N]$.

\begin{algorithm}
	\caption{Sampling Profile Scheme} \label{sampleprofile}
    \begin{algorithmic}[1]
    	\For {$i$ in $S$}
    		\For {$j=RowPtr[i],\ldots,RowPtr[i+1]$}
    			\For {$k$ in $\{4, 8, 16, 32\}$}
    				\State ColCounter[k][i][j/k] += 1
    			\EndFor
    			\State NnzElement[k][i] = RowPtr[i+1]-RowPtr[i]
    			\State NnzBitRow[k][i] = size of ColCounter[k][i]
    		\EndFor
    		\State EstCompressionRate[k] = avg(size of ColCounter[k][i] / size of ColCounter[k][i])
    	\EndFor
    \end{algorithmic} 
\end{algorithm}

The sampling result provides a rough estimation of the compression rate of Bit-GraphBLAS on B2SR-4 to B2SR-32. Users can select the affordable compression rate and tile size configuration.

\section{Bit Operations and BLAS Kernels Design}

This subsection discusses the code patterns and scheme designs that leverage the proposed storage format. Our implementation extensively uses the GPU integer intrinsics for bit-operations and other optimizations for computing efficiency, load balance, and memory performance.  

We briefly introduce the GPU bit operation intrinsics that will be used in this work, with respect to the BSTC bit-block abstraction \cite{BSTC}: (1) \textbf{\texttt{\_\_popc()}}: The population count function is for efficient bit-accumulation across a singe bit-row. CUDA supports population count along a 32-bit unsigned int. Paired with the logical AND operation, it can perform bit-dot-product for two 32-bit bit-rows. (2) \textbf{\texttt{\_\_shfl\_sync()}}: This intrinsic is for exchanging the bit-row across the lanes of a warp. For BMM, it facilitates a faster bit-dot-product between a bit-row and multiple bit-columns. (3) \textbf{\texttt{\_\_ballot\_sync()}}: This warp-vote intrinsic returns a 32-bit unsigned integer whose Nth bit indicates a predicate setting by the Nth thread of a warp (assuming all threads are active). This is essentially equal to transposing a bit-column to a bit-row. Since bits are indexed from right to left in the bit-row packing, the function is equivalent to a 90$^{\circ}$ clockwise transposition to a bit-row. (4) \textbf{\texttt{\_\_brev()}}: This intrinsic is used in bit-packing. Together with \textbf{\texttt{\_\_ballot\_sync()}}, this function rotates a bit-column 90$^{\circ}$ anti-clockwise into a bit-row.

The kernels of matrix-centric graph computing are matrix-vector and matrix-matrix computations. Table~\ref{bmv-scheme} and~\ref{bmm-scheme} list the core schemes we have implemented. They correspond to the different scenarios of the inputs and outputs (binary "1-bit" or full precision "32-bit"). We use two of the schemes to explain our implementations:

\begin{table}[]
\caption{BMV schemes used in the algorithms.}
\label{bmv-scheme}
\centering
\resizebox{0.9\columnwidth}{!}{
\begin{tabular}{|c|c|c|c|}
\hline
\textbf{Scheme}                         & \textbf{\begin{tabular}[c]{@{}c@{}}Input\\ Matrix A\end{tabular}} & \textbf{\begin{tabular}[c]{@{}c@{}}Input\\ Vec. B\end{tabular}} & \textbf{\begin{tabular}[c]{@{}c@{}}Output\\ Vec. C\end{tabular}} \\ \hline
\textbf{\texttt{bmv\_bin\_bin\_bin()}}           & 1-bit                                                           & 1-bit                                                           & 1-bit                                                            \\ \hline
\textbf{\texttt{bmv\_bin\_bin\_full()}}          & 1-bit                                                           & 1-bit                                                           & 32-bit                                                           \\ \hline
\textbf{\texttt{bmv\_bin\_full\_full()}}         & 1-bit                                                           & 32-bit                                                          & 32-bit                                                           \\ \hline
\textbf{\texttt{bmv\_bin\_bin\_bin\_masked()}}   & 1-bit                                                           & 1-bit                                                           & 1-bit                                                            \\ \hline
\textbf{\texttt{bmv\_bin\_bin\_full\_masked()}}  & 1-bit                                                           & 1-bit                                                           & 32-bit                                                           \\ \hline
\textbf{\texttt{bmv\_bin\_full\_full\_masked()}} & 1-bit                                                           & 32-bit                                                          & 32-bit                                                           \\ \hline
\end{tabular}
}
\end{table}

\textbf{Listing~\ref{listing-bmv}.} As an example scheme of Binarized Sparse Matrix Multiply Vector (BMV), Listing~\ref{listing-bmv} shows the code of \textbf{\texttt{bmv\_bin\_bin\_full()}} in 32$\times$32 tile size. The function demonstrates the bit multiplication between the binarized bit-tiles and the binarized vector. The result is a full-precision vector. The computation is as follows:
\[A^{(b)}_{i, j} \times b^{(b)}_{j} = c_{i} = \textbf{\texttt{\_\_popc($A^{(b)}_{i, j} \: \& \: b^{(b)}_{j}$)}}\]
Before computation, the sparse matrix is packed into the hierarchical storage format --- the vector is binarized into the column-major order with 32 consecutive elements compacted as an unsigned int. This allows the bit-columns to be fetched according to the same indexing system and enables fast bit-dot-product with each bit-row in the bit-tiles. The \textbf{\texttt{A}} variable is a bit-row in a tile, and the \textbf{\texttt{B}} variable is the binarized vector. The output array \textbf{\texttt{C}} is a vector in full precision. In each warp of the thread block, the number of bit-tiles to be computed is indicated by the \textit{TileRowptr}. In each iteration, the bit-tile and the 32-binarized vector perform bit-matrix-vector-product using bit-wise \textbf{\texttt{AND}} and \textbf{\texttt{\_\_popc()}}. Each lane is responsible for the output of a bit-row, whereas the registers being private to each thread accommodates the intermediate result per bit-tile. Finally, the content of the register is stored in the corresponding row of the resulting vector.

\begin{listing}[ht]
\begin{minted}
[
frame=lines,
framesep=2mm,
baselinestretch=1,
bgcolor=LightGray,
fontsize=\scriptsize
]{cpp}
int row_start = rowptr[bx];
int row_end = rowptr[bx+1];
if (row_start != row_end) {
    const unsigned* Asub = &(A[row_start*32]);
    const unsigned* Bsub = &(B[0]);
    T* Csub = &(C[bx*32]);
    register unsigned Cm[1] = {0};
    for (int i=row_start; i<row_end; i++) {
        unsigned r0 = Asub[(i-row_start)*32+laneid];
        unsigned r1 = Bsub[(colind[i])];
        Cm[0] += __popc(r0 & r1);
    }
    Csub[laneid] += (Cm[0]);
}
\end{minted}
\caption{BMV. A tile row is computed in a warp.}
\label{listing-bmv}
\end{listing}

\begin{table}[]
\caption{BMM schemes used in the algorithms.}
\label{bmm-scheme}
\centering
\resizebox{0.9\columnwidth}{!}{
\begin{tabular}{|c|c|c|c|}
\hline
\textbf{Schemes}                      & \textbf{\begin{tabular}[c]{@{}c@{}}Input\\ Matrix A\end{tabular}} & \textbf{\begin{tabular}[c]{@{}c@{}}Input\\ Mat. B\end{tabular}} & \textbf{\begin{tabular}[c]{@{}c@{}}Output\\ Value\end{tabular}} \\ \hline
\textbf{\texttt{bmm\_bin\_bin\_sum()}}         & 1-bit                                                           & 1-bit                                                           & 32-bit                                                          \\ \hline
\textbf{\texttt{bmm\_bin\_bin\_sum\_masked()}} & 1-bit                                                           & 1-bit                                                           & 32-bit                                                          \\ \hline
\end{tabular}
}
\end{table}

\textbf{Listing~\ref{listing-bmm}.} As an example of Binarized Sparse Matrix Multiply Binarized Sparse Matrix (BMM), Listing~\ref{listing-bmm} shows the code of \textbf{\texttt{bmm\_bin\_bin\_sum()}} for B2SR-32. The \textit{A} and \textit{B} variables are the two bit-vector in tiles of the input sparse matrices. The output \textbf{\texttt{C}} is a single variable in full precision, summing up the nonzeros (1s) of the resulting bit matrix. In each warp of the thread block, the number of bit-tiles in \textbf{\texttt{A}}'s tile row is indicated by the \textit{TileRowptr} of \textbf{\texttt{A}}. In the outer "for" loop, the \textit{TileColInd} of each bit-tile is used to retrieve the corresponding tile rows and bit-tiles of these tile-rows. The inner "for" loop performs the bit-tile-matrix-matrix-multiplication, where \textbf{\texttt{\_\_shfl\_sync()}} is used to retrieve the \textbf{\texttt{B}}'s bit-vectors in each lane. The temporary result of each bit-vector in \textbf{\texttt{B}} is accumulated in separate registers for avoiding race conditions. Finally, the content in the 32 registers is summed up. The sum is atomically added to a single destination element in \textbf{\texttt{C}}.

\begin{listing}[ht]
\begin{minted}
[
frame=lines,
framesep=2mm,
baselinestretch=1,
bgcolor=LightGray,
fontsize=\scriptsize
]{cpp}
T* Csub = &C[0];
register int Cm[32] = {0};
int sum = 0;
int A_row_start = A_rowptr[bx];
int A_row_end = A_rowptr[bx+1];
const unsigned* Asub = &(A[A_row_start*32]);
for (int i=A_row_start; i<A_row_end; i++) {
    unsigned r0 = Asub[(i-A_row_start)*32+laneid];
    int A_col = A_colind[i];
    int B_row_start = B_rowptr[A_col];
    int B_row_end = B_rowptr[A_col+1];
    const unsigned* Bsub = &(B[B_row_start*32]);
    for (int j=B_row_start; j<B_row_end; j++) {
        unsigned r1 = Bsub[(j-B_row_start)*32
                      +laneid];
        #pragma unroll
        for (int k=0; k<32; k++){
            unsigned r2 = __shfl_sync(0xFFFFFFFF, 
                          r1, k);
            Cm[k] += __popc(r0 & r2);
        }
}
\end{minted}
\caption{BMM. A tile row is computed in a warp.}
\label{listing-bmm}
\end{listing}

\begin{figure}
    \centering
    \includegraphics[trim={0cm 0cm 7cm 0cm},clip,width=0.5\textwidth]{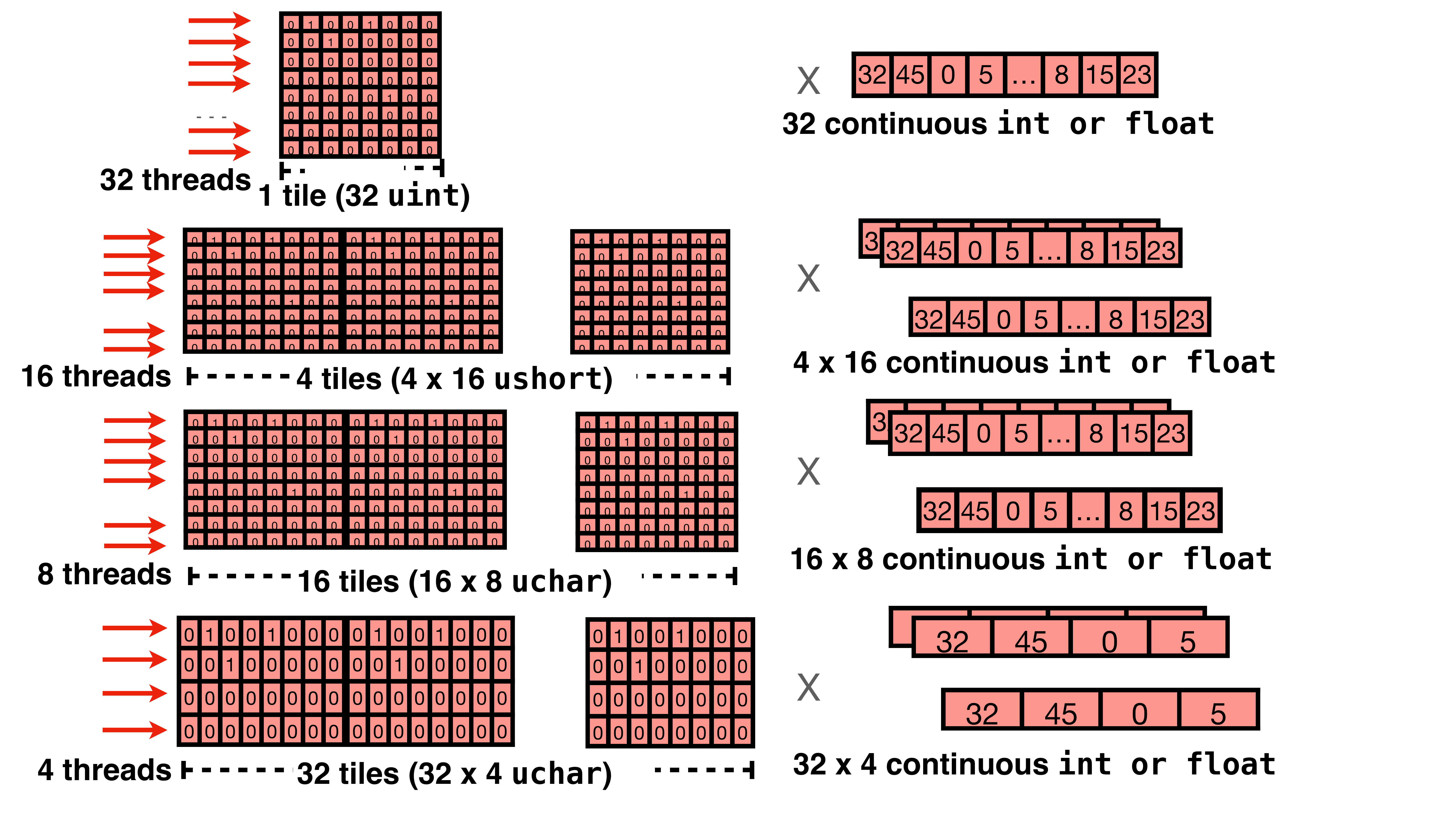}
    \caption{Illustration of \textbf{\texttt{bmv\_bin\_full\_full()}}.}
    \label{fig-fullvector}
\end{figure}

\textbf{Allowing efficient full-precision vector load}
In BMV and BMM, the bit-tiles in a tile-row are handled in a warp of 32 threads, following the warp-consolidation model~\cite{li2018warp}. By default, each of the thread blocks contains only one warp. So up to 64 thread blocks can be freely scheduled by a single SM scheduler. This works fine when kernels are computing in a binarized vector or matrix. However, in \textbf{\texttt{bmv\_bin\_full\_full()}}, when engaging a full-precision vector as the multiplier, the one-warp-per-thread-block design impedes the flexibility to preload the common vector portions for the neighboring tile rows into shared memory. Thus, we implement the kernel scheme with 32 warps processing consecutive 32 tile-rows in a thread block. This further enhances the spatial locality of the computation workload. Figure~\ref{fig-fullvector} shows the thread mapping for B2SR-32, B2SR-16, B2SR-8, B2SR-4. The execution latency is generally shorter when cooperatively loading 128 bytes (equal to the cache line size of GPUs in our experiment) of binarized tiles from global memory. At the same time, the corresponding number of full-precision subvectors have to be loaded for multiplication. We set the thread block to contain 1024 threads to load the vectors into shared memory before multiplication. Bit-vectors on tile-row with the same column index can share the preload vectors from shared memory.

\section {Graph Algorithms} \label{graph-algo-section} 
\begin{table}[]
\caption{Semiring support with BMV and BMM schemes.}
\label{tab-semiring}
\centering
\resizebox{\columnwidth}{!}{
\begin{tabular}{|c|c|c|c|}
\hline
\textbf{Semiring}  & \textbf{Domain}                & \textbf{Algorithm}                                                    & \textbf{Scheme}                                                                             \\ \hline
Boolean            & $\{0, 1\}$                     & \begin{tabular}[c]{@{}c@{}}BFS, diameter, \\ MIS, GC\end{tabular} & \textbf{\texttt{bin-bin-bin}}                                                                        \\ \hline
Arithmetic         & $\mathbf{R}$                   & LGC, PR, TC                                                       & \begin{tabular}[c]{@{}c@{}}\textbf{\texttt{bin-full-full}}\\ or\\ \textbf{\texttt{bin-bin-full}}\end{tabular} \\ \hline
Tropical Min-plus  & $\mathbf{R} \cup \{+ \infty\}$ & SSSP, CC                                                          & \textbf{\texttt{bin-full-full}}                                                                      \\ \hline
Tropical Max-times & $\mathbf{R}$                   & MIS, GC                                                           & \textbf{\texttt{bin-full-full}}                                                                      \\ \hline
\end{tabular}
}
\end{table}

This section explains how graph programs can be implemented upon the core operations. 

Matrix-centric graph computing models graph traversals as operations on semirings~\cite{kepner2011graph}. 
As shown in Table~\ref{tab-semiring}, our implementation can support the key semiring domain operations when performing \textbf{\texttt{vxm()}}, \textbf{\texttt{mxv()}}, \textbf{\texttt{mxm()}}. After the adjacency matrix is in B2SR, it remains binary throughout all operations. The vectors representing the frontier nodes are all in dense format. They can be either binarized for binary semiring or full-precision (float, unsigned, bool, etc.) for the non-binary domains to support a variety of graph algorithms. We also implement efficient masking schemes for both BMV and BMM: \textbf{\texttt{bmv\_bin\_bin\_bin\_masked()}}, \textbf{\texttt{bmv\_bin\_bin\_full\_masked()}}, \textbf{\texttt{bmv\_bin\_full\_\\full\_masked()}}, and \textbf{\texttt{bmm\_bin\_bin\_sum\_masked()}}. We next use two graph algorithms to illustrate how to utilize these kernel backends when writing graph algorithms.

\textbf{Breadth-First-Search (BFS)}
Breadth-first-search uses boolean semiring. In each iteration, the \textbf{\texttt{vxm()}} performs one-degree edge traversal to all the connected vertices. A mask of visited vertices is applied at the end to filter out the visited results. We introduce \textbf{\texttt{bmv\_bin\_bin\_bin\_masked()}} to enable this. GraphBLAST uses early exit to eliminate the masked element operations in their masked \textbf{\texttt{vxm()}}. Yet, a similar strategy does not apply to our case. In our implementation, the consecutive rows in a tile row are operated in the same warp. Early exit causes a performance penalty because of warp divergence. Therefore, in the \textbf{\texttt{bmv\_bin\_bin\_bin\_masked()}} kernel, the bitmask is applied right before the output store, having bit-wise \textbf{\texttt{AND}} with the negation of visited vertex vector (indicates unvisited vertices).

\textbf{Single-Source Shortest-Path (SSSP)}  We implement the algorithm with delta-stepping SSSP~\cite{davidson2014work} as in GraphBLAST. SSSP utilizes tropical min-plus semiring. The intermediate vectors are reduced by minimum operation. \textbf{\texttt{bmv\_bin\_full\_full()}} maintains the multiplier vector in full-precision, allowing it to carry minimum distance values. To realize the relaxation, we set an extra condition in the  \textbf{\texttt{bmv\_bin\_full\_full()}} such that the $0$s in the adjacency matrix are identified as infinite ($\infty$), indicating unreachable. Only $0$s along the diagonal are treated as actual zeros, which we omit their self-connectivity. Within a warp, a thread reduces all non-zero full-precision values along the multiplier vector by \textbf{\texttt{Min()}} (for B2SR-32). In B2SR-4, B2SR-8, and B2SR-16, since we use more than one thread to process the values along the multiplier vector, \textbf{\texttt{atomicMin()}} is applied to avoid race conditions. 

\textbf{PageRank (PR)} PR uses arithmetic semiring. In each iteration, the page rank vector is multiplied by the \textit{column stochastic adjacency matrix}. The \textit{column stochastic adjacency matrix} is the adjacency matrix with each out-vertex connectivity divided by the vertex's out-degree. Since the page rank vector is in full-precision, we use \textbf{\texttt{bmv\_bin\_full\_full()}} with an auxiliary vector \textbf{\texttt{v\_out\_degree}} to accommodate each vertex's out-degree. For each $1$ on the matrix, the corresponding value on the page rank vector is divided by its out-degree on \textbf{\texttt{v\_out\_degree}}. Eventually, the intermediate vector is summed up with add operation to the output, indicating the weighted sum. Likewise, B2SR-4, B2SR-8, and B2SR-16 requires \textbf{\texttt{atomicAdd()}} since more than one thread process the workload along the vector cooperatively.

\textbf{Connected Component (CC)} We follow the CC implementation in GraphBLAST, which is based on the FastSV linear-algebraic connected component alogrithm~\cite{zhang2020parallel, zhang2020fastsv}. Similar to SSSP, CC uses tropical min-plus semiring. We adopt \textbf{\texttt{bmv\_bin\_full\_full()}} since the frontier vector should be in full-precision. The \textbf{\texttt{mxv()}} is achieved by reducing the non-zero full-precision values along the intemediate vector using \textbf{\texttt{Min()}} and \textbf{\texttt{atomicMin()}}.

\textbf{Triangle Counting (TC)} We implement TC as in GraphBLAST, following Azas and Buluc's~\cite{tcazas} and Wolf's~\cite{wolf2017fast}. The TC uses arithmetic semiring. It is achieved by multiplying the lower triangle of the adjacency matrix ($L$) with the transpose of itself ($L^T$) and then applying ($L$) as the mask to generate the output matrix. Ultimately, the non-zeros is summed up into one full-precision value; therefore, we fuse the reduction sum kernel with \textbf{\texttt{mxm()}} and directly perform \textbf{\texttt{atomicAdd()}} to global sum once a bitmap subroutine finishes. Since both input matrices and mask matrix can be sufficiently represented in binary format, so we realize the kernel through \textbf{\texttt{bmm\_bin\_bin\_sum\_masked()}}.

\begin{table}[]
  \centering
  \resizebox{\columnwidth}{!}{
  \begin{tabular}{ | c | c | c | c |}
    \hline
    \textbf{Category} & \textbf{Example} & \textbf{\% in Dataset} &\textbf{Description}\\ \hline
    \textbf{Dot}
    &
    \begin{minipage}{.1\textwidth}
      \includegraphics[trim={2cm 2cm 2cm 2cm},clip,width=\linewidth]{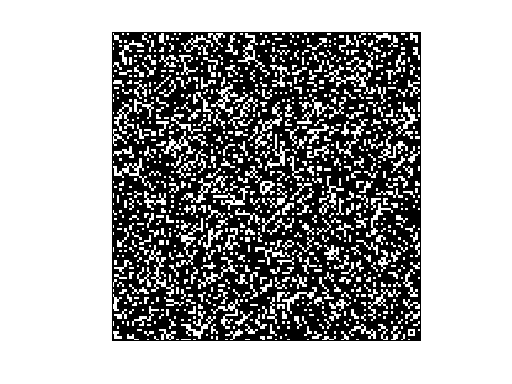}
    \end{minipage}
    &
    36.66\%
    &
    \begin{tabular}{l}
    \begin{minipage}[t]{2cm}
    Contains nonzeros scatter randomly
    \end{minipage}
    \end{tabular}
    \\ \hline
    \textbf{Diagonal}
    &
    \begin{minipage}{.1\textwidth}
      \includegraphics[trim={1cm 1cm 1cm 1cm},clip,width=\linewidth]{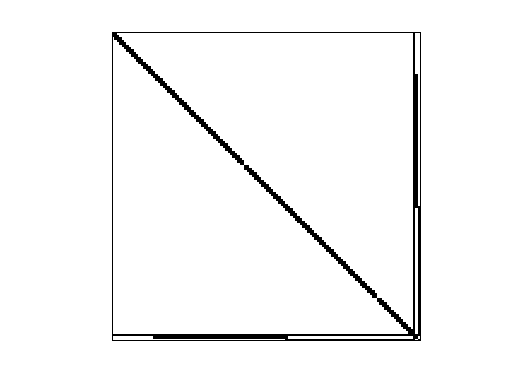}
    \end{minipage}
    &
    45.87\%
    &
    \begin{tabular}{l}
    \begin{minipage}[t]{2cm}
    Nonzeros are centralized around diagonal
    \end{minipage}
    \end{tabular}
    \\ \hline
    \textbf{Block}
    &
    \begin{minipage}{.1\textwidth}
      \includegraphics[trim={1cm 1cm 1cm, 1cm},clip,width=\linewidth]{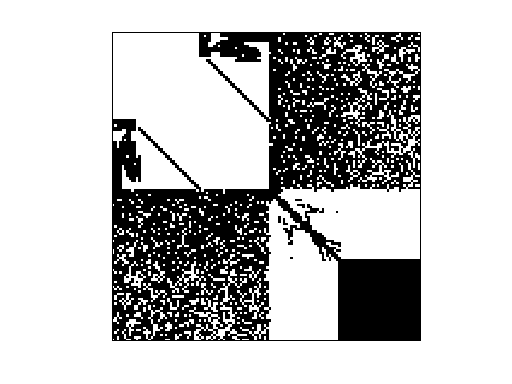}
    \end{minipage}
    &
    24.95\%
    &
    \begin{tabular}{l}
    \begin{minipage}[t]{2cm}
    Square or rectangular blocks, countours
    \end{minipage}
    \end{tabular}
    \\ \hline
    \textbf{Stripe}
    &
    \begin{minipage}{.1\textwidth}
      \includegraphics[trim={1cm 1cm 1cm 1cm},clip,width=\linewidth]{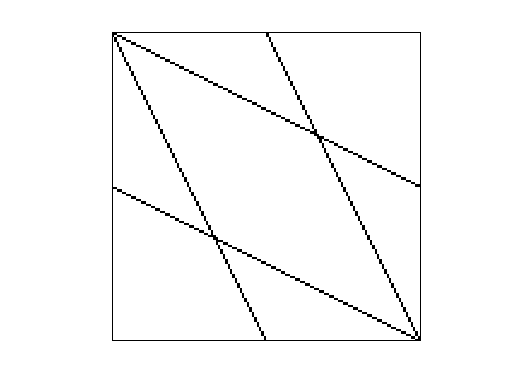}
    \end{minipage}
    &
    13.05\%
    &
    \begin{tabular}{l}
    \begin{minipage}[t]{2cm}
    Contain one or more lines in various directions
    \end{minipage}
    \end{tabular}
    \\ \hline
    \textbf{Road}
    &
    \begin{minipage}{.1\textwidth}
      \includegraphics[trim={1cm 1cm 1cm 1cm},clip,width=\linewidth]{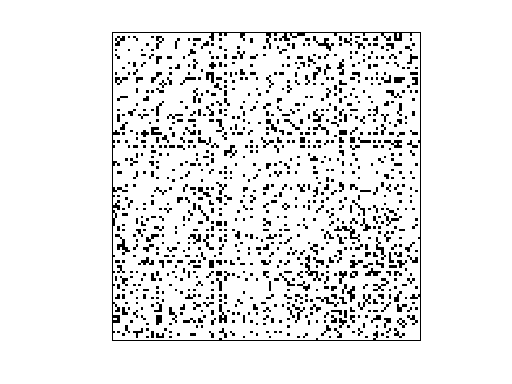}
    \end{minipage}
    &
    5.18\%
    &
    \begin{tabular}{l}
    \begin{minipage}[t]{2cm}
    Nonzeros in regular distribution
    \end{minipage}
    \end{tabular}
    \\ \hline
    \textbf{Hybrid}
    &
    \begin{minipage}{.1\textwidth}
      \includegraphics[trim={1cm 1cm 1cm 1cm},clip,width=\linewidth]{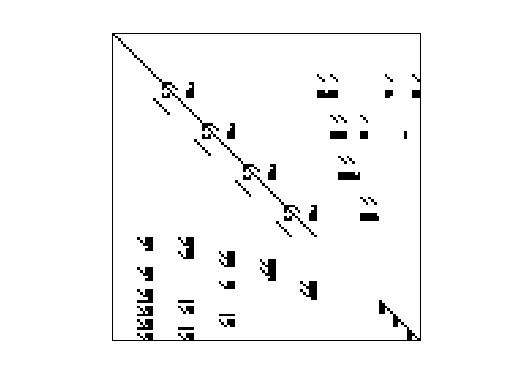}
    \end{minipage}
    &
    25.72\%
    &
    \begin{tabular}{l}
    \begin{minipage}[t]{2cm}
    A combination of more than two patterns above
    \end{minipage}
    \end{tabular}
    \\ \hline
  \end{tabular}
  }
  \caption{Matrix pattern category (black indicates nonzeros).}\label{mat-pattern}
\end{table}

\section{Evaluation}
\label{sec:eval}

\subsection{Experiment Configuration}
\textbf{Dataset}
We use all 521 binary square matrices in the SuiteSparse Matrix Collection~\cite{SuiteSparse}. The set of matrices contains the number of rows and columns ranging from 2 to 214,005,017 and the number of nonzeros from 2 to 11,588,725,964. To better summarize the similarity between the matrices with higher or lower performance in the evaluation, we further classify the matrices into six categories based on their patterns in Table~\ref{mat-pattern}. 

\begin{table}[]
\caption{GPU memory and cache hierarchy.}
\label{tab-gpu-memory}
\resizebox{\columnwidth}{!}{
\begin{tabular}{|c|c|c|c|c|c|c|c|c|}
\hline
\textbf{GPU} & \textbf{Arch} & \textbf{SMs} & \textbf{\begin{tabular}[c]{@{}c@{}}Shared/\\ SM\end{tabular}} & \textbf{\begin{tabular}[c]{@{}c@{}}Shared/\\ Block\end{tabular}} & \textbf{RAM} & \textbf{\begin{tabular}[c]{@{}c@{}}Memory \\ Bandwidth\end{tabular}} & \textbf{\begin{tabular}[c]{@{}c@{}}L1 Cache \\ Size/SM\end{tabular}} & \textbf{\begin{tabular}[c]{@{}c@{}}L2 Cache \\ Size\end{tabular}} \\ \hline
GTX1080      & Pascal        & 20           & 64KB                                                          & 48KB                                                             & 8GB          & 320GB/sec                                                            & 48KB                                                                 & 2048KB                                                            \\ \hline
TITAN V      & Volta         & 80           & 96KB                                                          & 96KB                                                             & 12GB         & 653GB/sec                                                            & 96KB                                                                 & 4608KB                                                            \\ \hline
\end{tabular}
}
\end{table}

\textbf{GPU Environments} We evaluate the proposed format and computation core functions on two NVIDIA GPU architectures, including Pascal and Volta. With the compute capability 6.0 and 7.0 configured, respectively, how the proposed format adapts to each hardware-specific variance is worth seeing. Table~\ref{tab-gpu-memory} shows the configured SM information and memory hierarchy size of the two GPU architectures in the evaluation. We use CUDA version 10.0 across all our evaluations.

\textbf{Algorithm Parameters}
The optimization setup of GraphBLAST's algorithms is based on the default running script provided in their GitHub repository~\cite{graphblastgithub}. BFS is with early-exit, structure-only, and operand reuse enabled. PR is limited to a maximum iteration of 10. The alpha parameter is set to 0.85, and pdfilon is set to $1e-9$. The runtime for all algorithms is measured by the average of 5 runs.

\angcmtfix{Where is the place that the graph pattern is discussed for the dataset? E.g., how the dataset is selected?}


\begin{figure}
    \begin{subfigure}[b]{.24\textwidth}
    \centering
    \includegraphics[scale=.24,width=\textwidth]{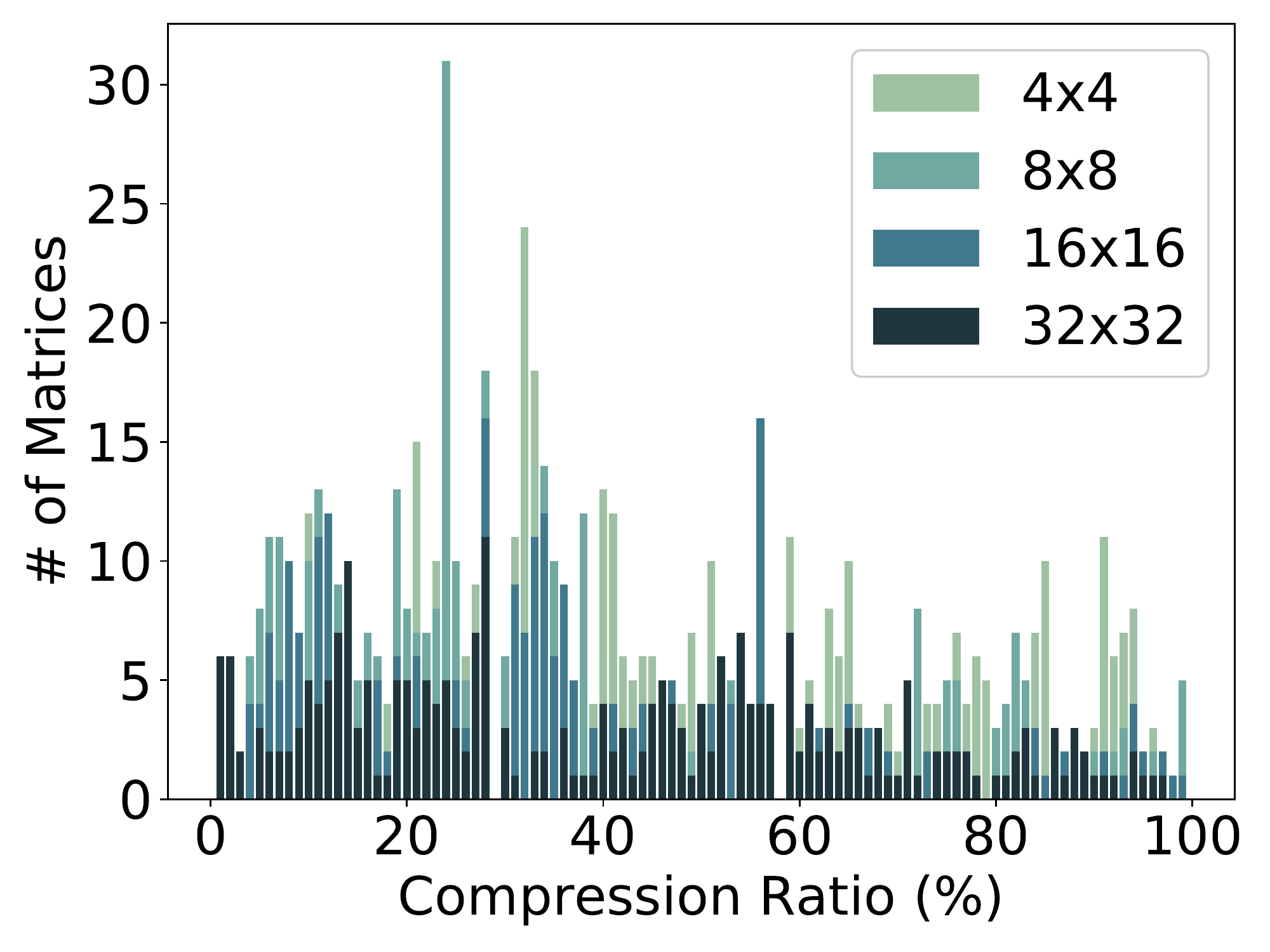}
    \caption{Compression rate histogram.}\label{fig-compression-histogram}
    \end{subfigure}
    \begin{subfigure}[b]{.24\textwidth}
    \centering
    \includegraphics[scale=.24,width=\textwidth]{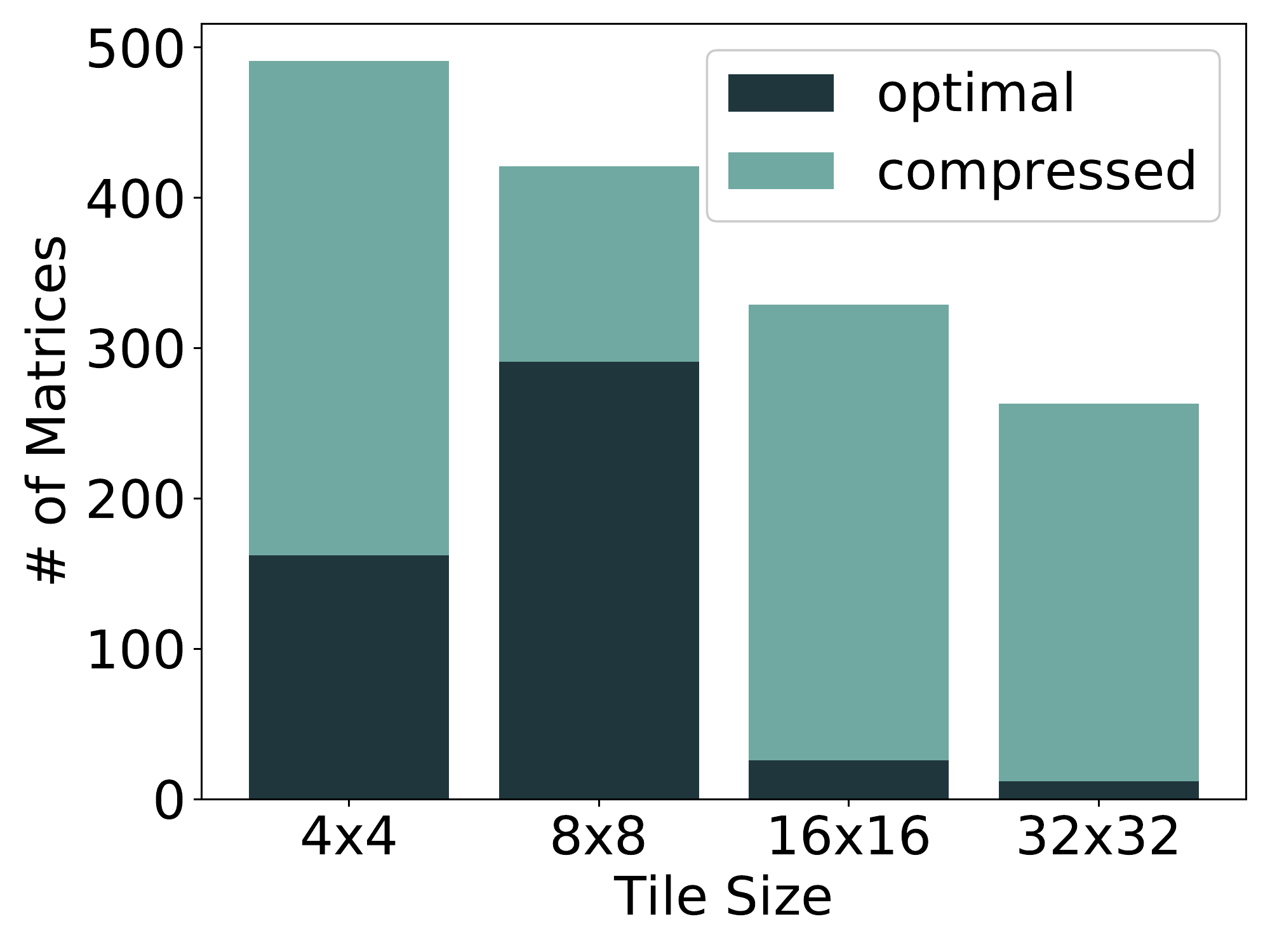}
    \caption{Optimal compression tile size.}\label{fig-optimal-compression}
    \end{subfigure}
    \caption{Compression result of the 521 binary matrices.}\label{compression}
\end{figure}

\subsection{Storage Efficiency} 
B2SR brings significant storage savings for large sparse matrices. We show the compression ratio of the 521 binary graph matrices with respect to the default 32-bit floating-point CSR. The compression ratio thus is defined as: $\frac{B2SR\_size}{CSR\_size}$. A lower value indicates a better compression rate. The compression ratio depends mainly on the nonzero distribution of the binary matrices. Figure~\ref{fig-compression-histogram} shows the compression ratio on the x-axis and the histogram recording the number of matrices using the four B2SR formats on the y-axis. In Figure~\ref{fig-optimal-compression}, the y-axis shows the number of matrices that belong to their: (1) optimal size (colored in blue): the least storage size required among the four B2SR formats of a matrix. (2) compressed size (colored in green): the B2SR format can provide a compression ratio $< 100\%$ for a matrix. For optimal, 162 matrices appear at B2SR-4, 291 matrices at B2SR-8, 26 matrices at B2SR-16, and 12 matrices at B2SR-32. For compressed, 491 matrices can have a compression ratio $< 100\%$ on B2SR-4, 421 on B2SR-8, 329 on B2SR-16, and 263 on B2SR-32.

\subsection{Overview of the Performance Gains}
There are multiple factors that contributed to the significant speedups B2SR has achieved. In addition to the gains by using native bit-level intrinsics such as \textbf{\texttt{\_\_popc()}}, we have observed that more performance is from the reduced memory transactions and enhanced data locality. For example, for the matrix \textit{mycielskian8}, by using B2SR, the number of global memory load transactions reduces by 4$\times$ from 6630 to 1826, while the L1 cache hit-rate increases by 24\% from 65.63\% to 81.83\%. We also observed different sweet areas for different B2SR tile sizes---such as, the smaller tile sizes (e.g., 4, 8) draw better L1 hit rates while the larger tile sizes (e.g., 32, 64) favor coalesced memory access---and also the impact from the profiles of individual graphs. In the next two sections, we describe the performance evaluation of the linear algebra kernels (BMV and BMM) and the five graph algorithms, respectively. 

\subsection{Linear Algebra Kernels} \label{subsection-kernels}
\begin{figure*}
    \begin{subfigure}[b]{.25\textwidth}
    \centering
    \includegraphics[scale=.25, width=\textwidth]{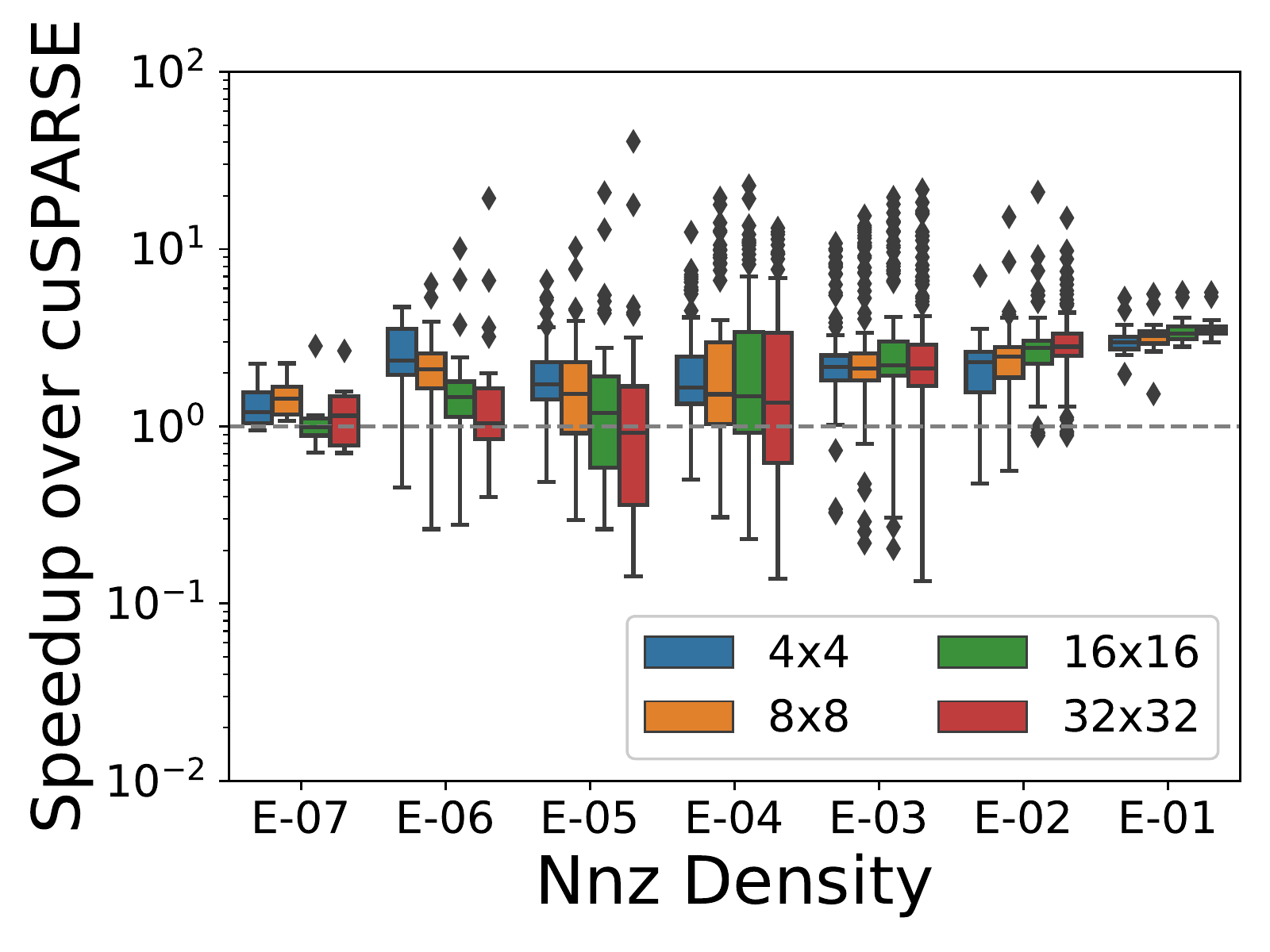}
    \caption{\textbf{\texttt{bmv\_bin\_bin\_bin()}}.} \label{bmv-bin-bin-bin-pascal}
    \end{subfigure}\hfill
    \begin{subfigure}[b]{.25\textwidth}
    \centering
    \includegraphics[scale=.25, width=\textwidth]{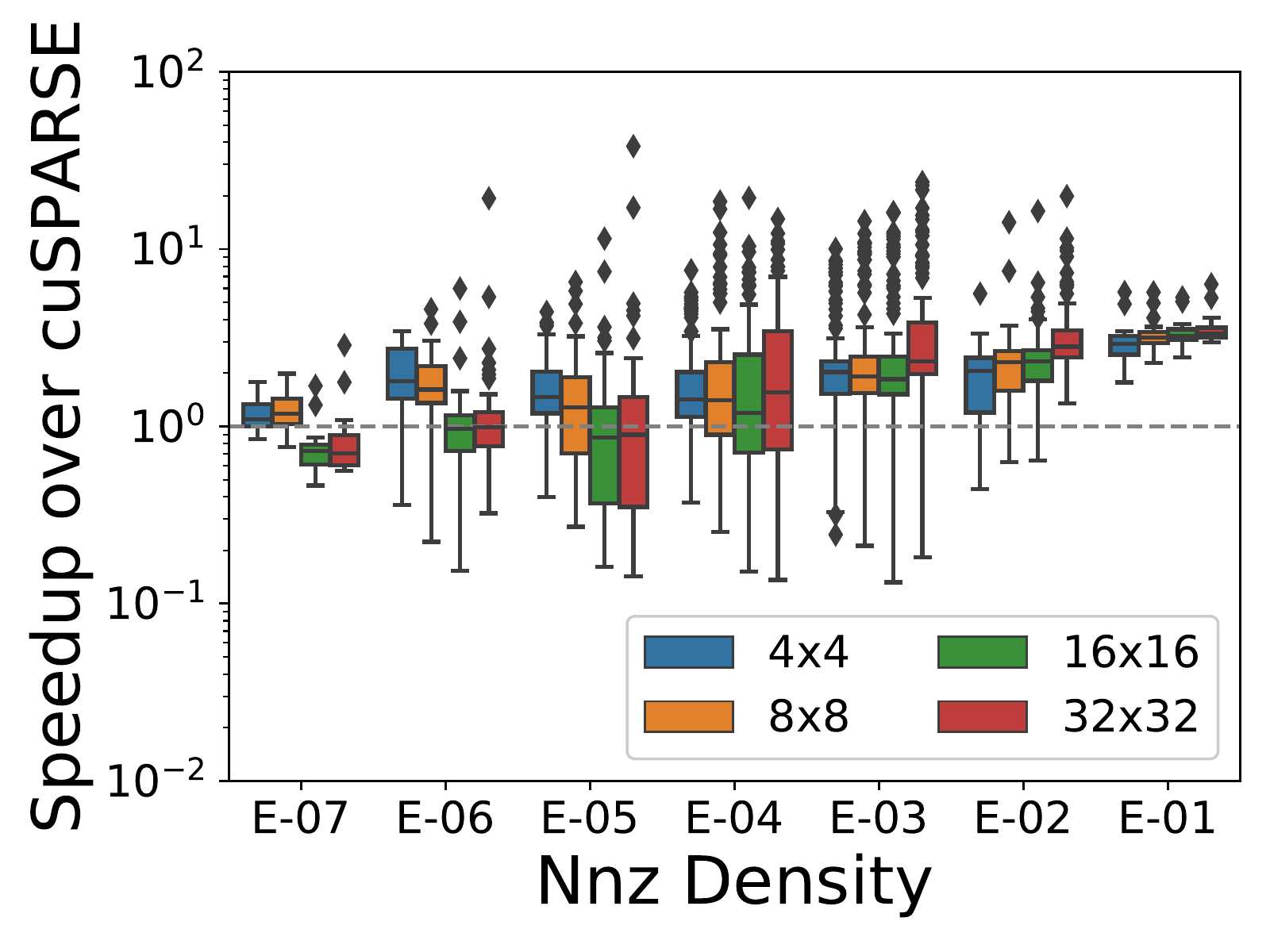}
    \caption{\textbf{\texttt{bmv\_bin\_bin\_full()}}.} \label{bmv-bin-bin-full-pascal}
    \end{subfigure}\hfill
    \begin{subfigure}[b]{.25\textwidth}
    \centering
    \includegraphics[scale=.25, width=\textwidth]{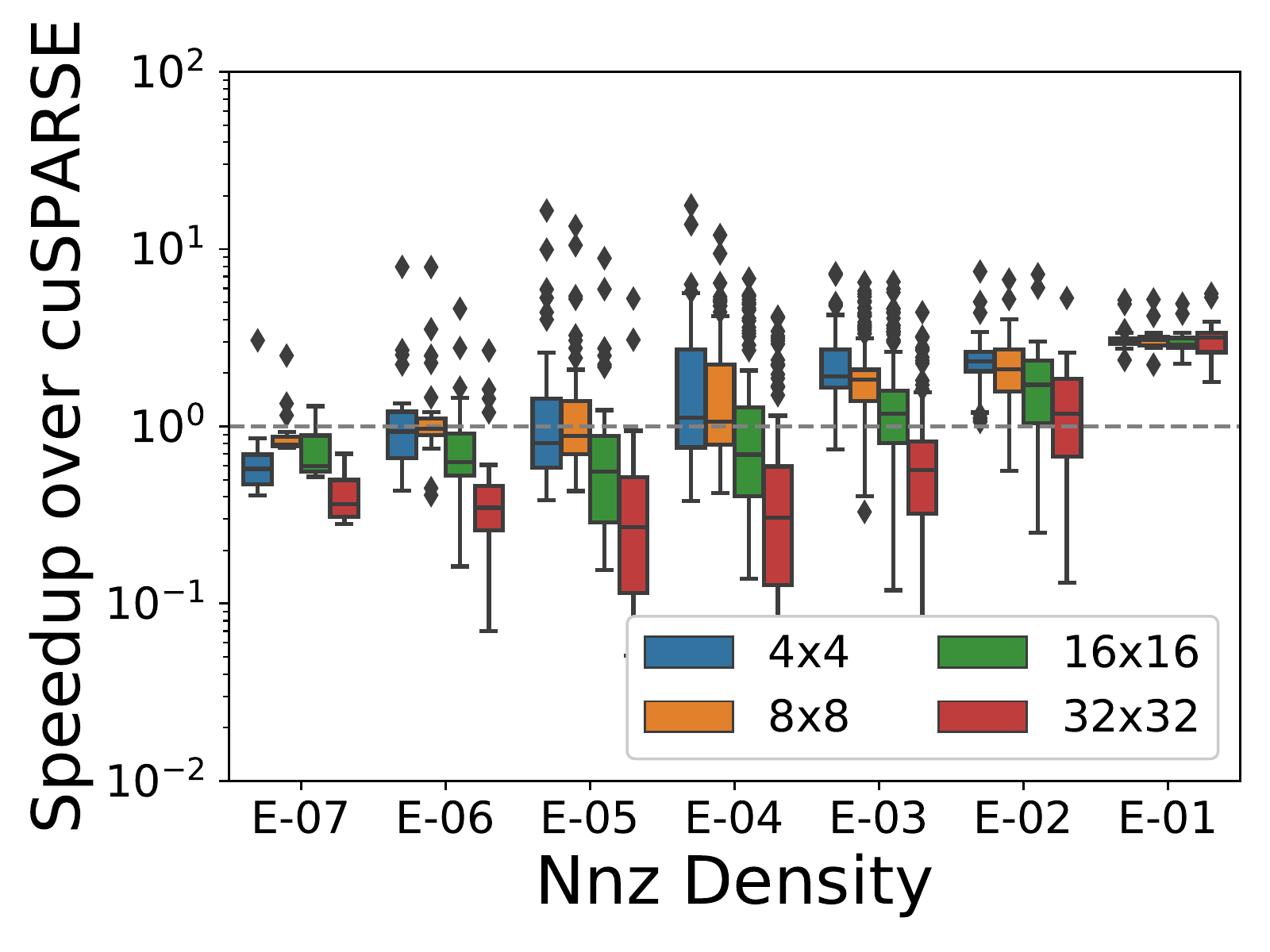}
    \caption{\textbf{\texttt{bmv\_bin\_full\_full()}}.} \label{bmv-bin-full-full-pascal}
    \end{subfigure}\hfill
    \begin{subfigure}[b]{.25\textwidth}
    \centering
    \includegraphics[scale=.25, width=\textwidth]{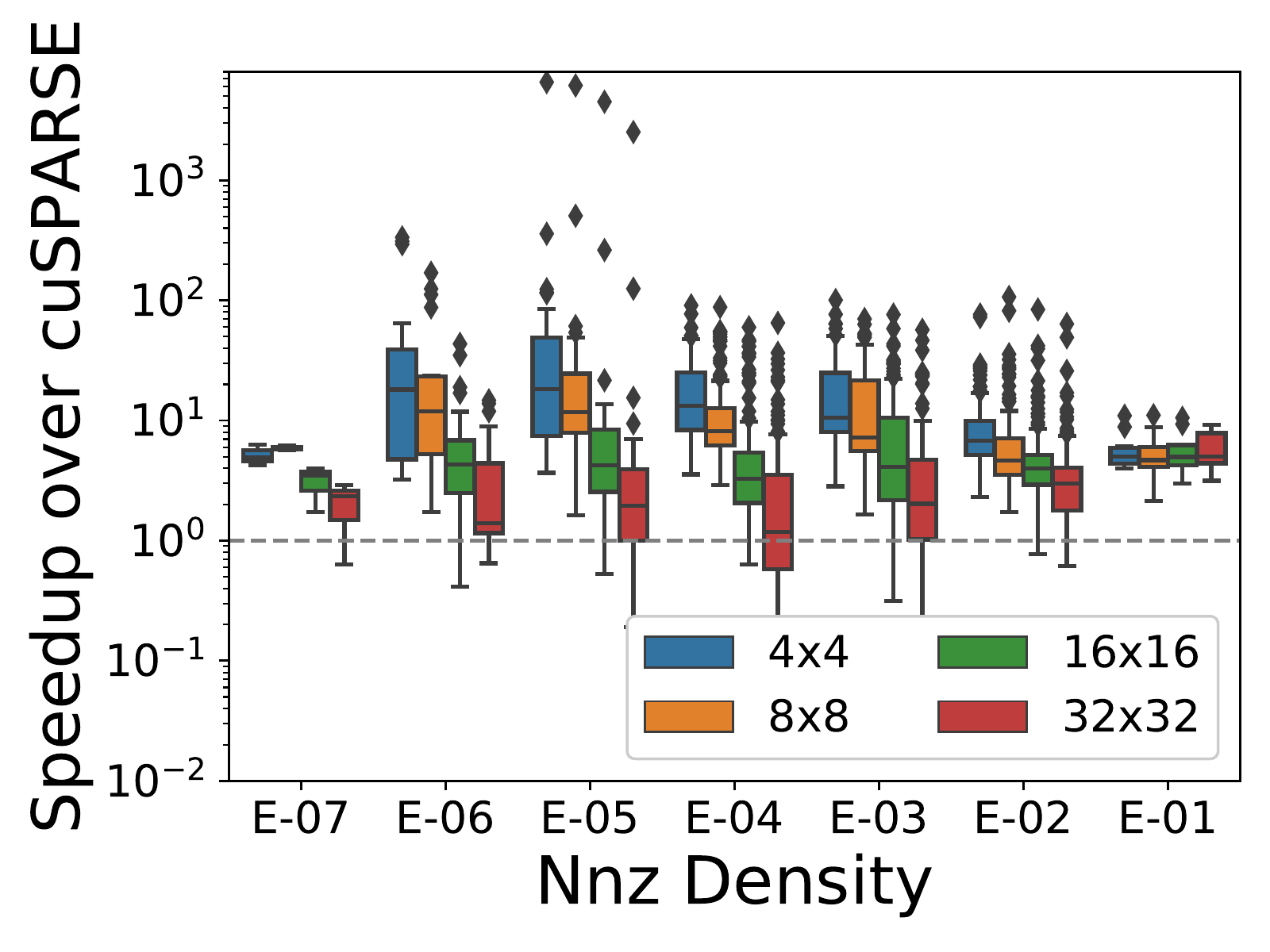}
    \caption{\textbf{\texttt{bmm\_bin\_bin\_sum()}}.} \label{bmm-bin-bin-sum-pascal}
    \end{subfigure}
    \caption{Arithmetic kernel speedup over CuSPARSE on GTX1080 (Pascal) GPU.}\label{fig-pascal}
\end{figure*}

\begin{figure*}
    \begin{subfigure}[b]{.25\textwidth}
    \centering
    \includegraphics[scale=.25, width=\textwidth]{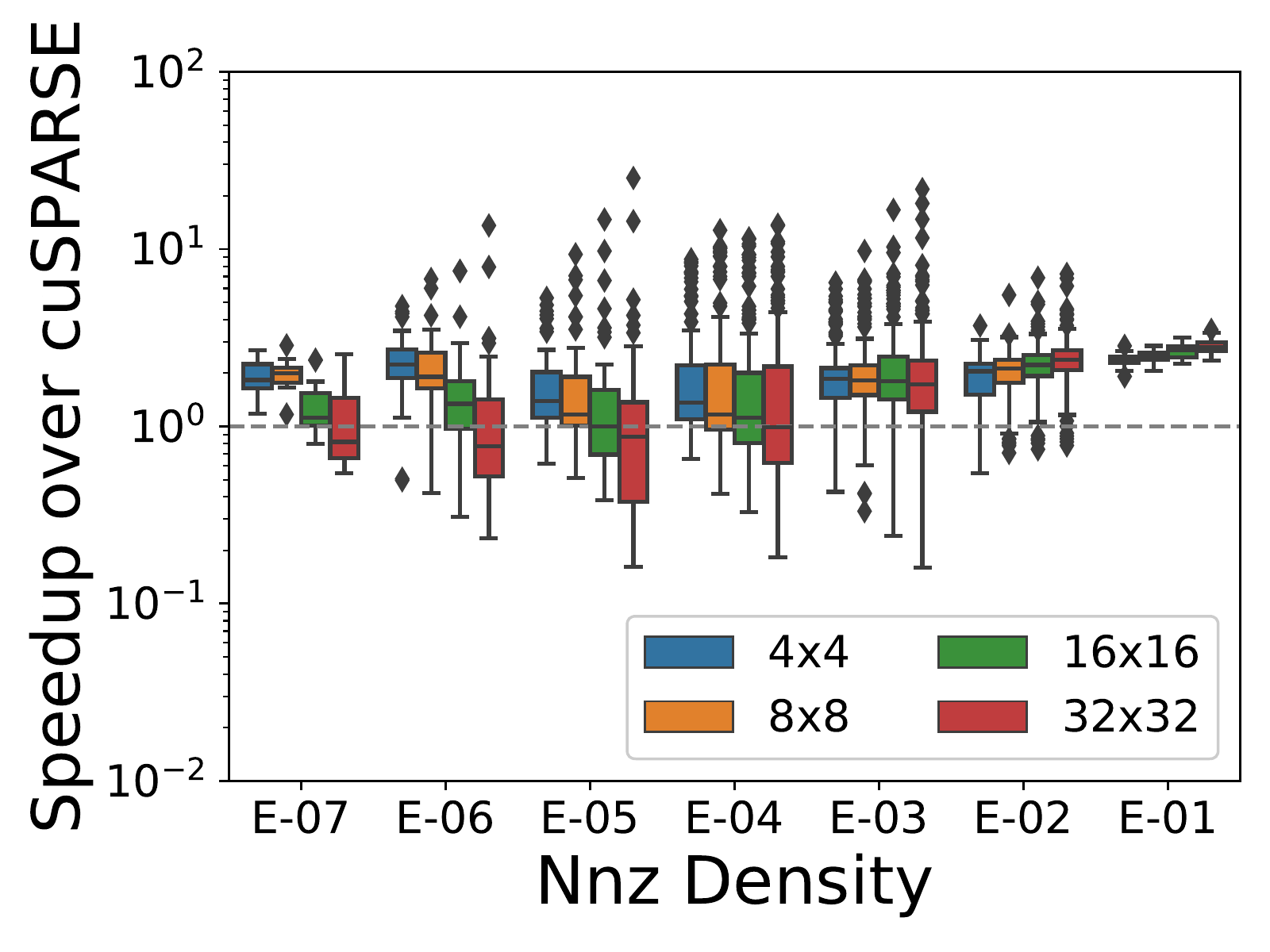}
    \caption{\textbf{\texttt{bmv\_bin\_bin\_bin()}}.} \label{bmv-bin-bin-bin-volta}
    \end{subfigure}\hfill
    \begin{subfigure}[b]{.25\textwidth}
    \centering
    \includegraphics[scale=.25, width=\textwidth]{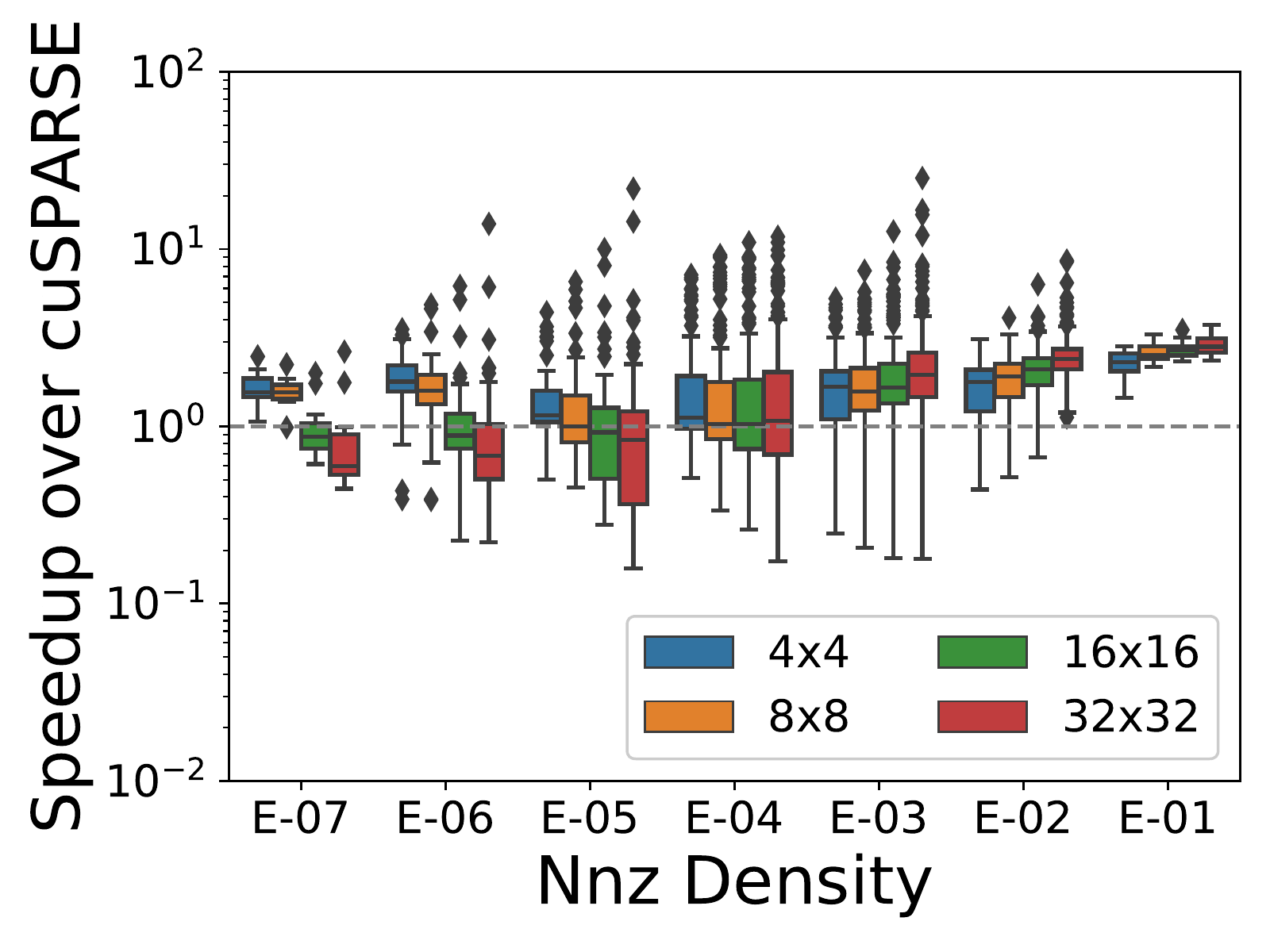}
    \caption{\textbf{\texttt{bmv\_bin\_bin\_full()}}.} \label{bmv-bin-bin-full-volta}
    \end{subfigure}\hfill
    \begin{subfigure}[b]{.25\textwidth}
    \centering
    \includegraphics[scale=.25, width=\textwidth]{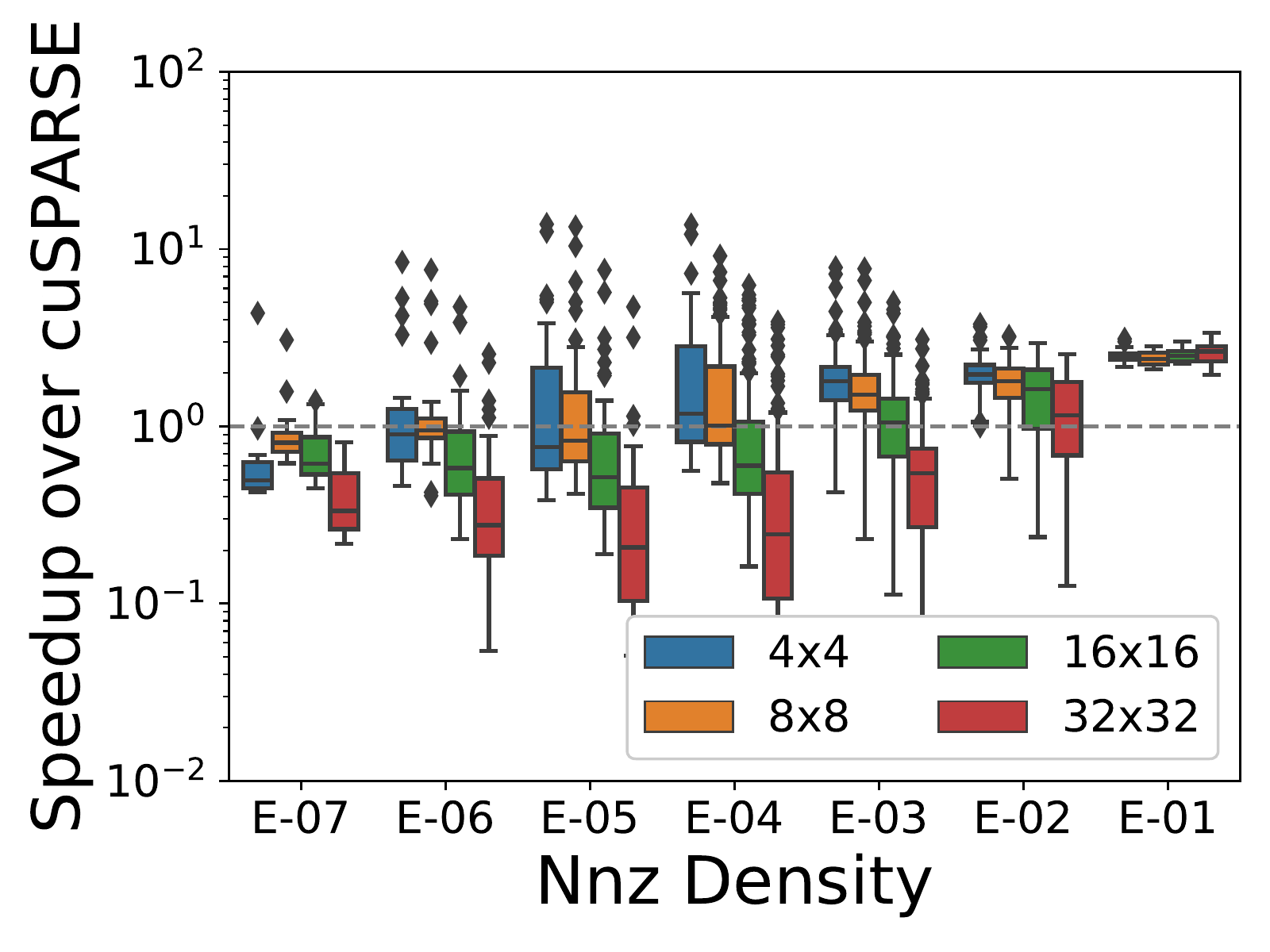}
    \caption{\textbf{\texttt{bmv\_bin\_full\_full()}}.} \label{bmv-bin-full-full-volta}
    \end{subfigure}\hfill
    \begin{subfigure}[b]{.25\textwidth}
    \centering
    \includegraphics[scale=.25, width=\textwidth]{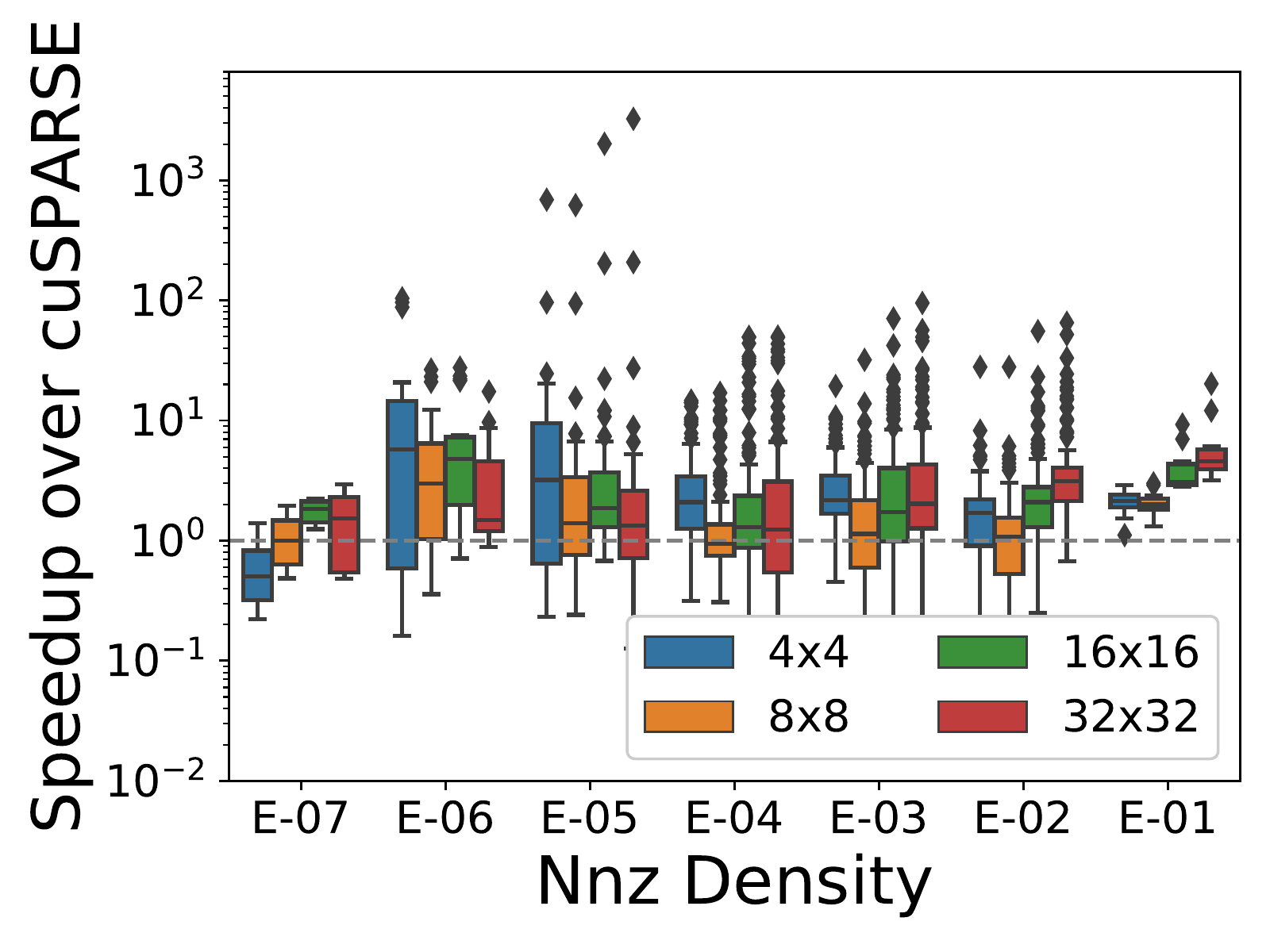}
    \caption{\textbf{\texttt{bmm\_bin\_bin\_sum()}}.} \label{bmm-bin-bin-sum-volta}
    \end{subfigure}
    \caption{Arithmetic kernel speedup over CuSPARSE on Titan V (Volta) GPU.}\label{fig-volta}
\end{figure*}

In this subsection, we evaluate the basic arithmetic cores BMV and BMM in terms of different schemes. We evaluate the speedups of the kernels over cuSPARSE's SpMV (\textbf{\texttt{cusparseScsrmv()}}) and SpGEMM (\textbf{\texttt{cusparseScsrgemm()}}) with CSR in 32-bit floating-point nonzero storage. We compare the performance of each kernel scheme on B2SR-4, B2SR-8, B2SR-16, and B2SR-32. In Figure~\ref{fig-pascal} and~\ref{fig-volta}, the y-axis is the average speedups (of 5 runs) over cuSPARSE and the x-axis is the nonzero density of the matrices which is defined as $\frac{\#\_of\_nonzeros}{\#\_of\_elements}$. A higher nonzero density (more to the right in the figure) implies a denser matrix while a lower one (more to the left) implies a sparser matrix.

\textbf{BMV} In BMV, we implement three schemes as the basic SpMV kernels (see Table~\ref{bmv-scheme}). We evaluate the performance of \textbf{\texttt{bmv\_bin\_bin\_bin()}}, \textbf{\texttt{bmv\_bin\_bin\_full()}},  \textbf{\texttt{bmv\_bin\_full\_full()}} and compare it with cuSPARSE's \textbf{\texttt{cusparseScsrmv(). bmv\_bin\_bin\_bin()}}'s performance is shown in Figure~\ref{bmv-bin-bin-bin-pascal} and~\ref{bmv-bin-bin-bin-volta}. Although the arithmetic capability of this scheme is limited to only binary operations, it allows a minimal memory footprint by keeping all value storage in binarized format. On GTX1080, \textbf{\texttt{bmv\_bin\_bin\_bin()}} achieves an average speedup of 2.41$\times$ in B2SR-4, 2.74$\times$ in B2SR-8, 2.91$\times$ in B2SR-16, and 2.85$\times$ in B2SR-32. The max speedup over baseline is 40.47$\times$ that appears at matrix \textit{ins2} with B2SR-32. On Titan V, \textbf{\texttt{bmv\_bin\_bin\_bin()}} achieves an average speedup of 2.04$\times$ in B2SR-4, 2.17$\times$ in B2SR-8, 2.18$\times$ in B2SR-16, and 2.26$\times$ in B2SR-32. The max speedup over baseline is 25.16$\times$ that appears at matrix \textit{ins2} with B2SR-32. 

\begin{table*}[!ht]
\caption{SpMV-based graph algorithm performance on GTX1080 (Pascal) GPU.}
\label{tab-spmv-algo-pascal}
\centering
\resizebox{1.6\columnwidth}{!}{
\begin{tabular}{|c|c|c|c|c|c|c|c|c|c|c|c|c|c|}
\hline
\multirow{2}{*}{\textbf{Matrix}}            & \textbf{Algorithm}                                                   & \multicolumn{3}{c|}{\textbf{BFS}}                         & \multicolumn{3}{c|}{\textbf{SSSP}}                      & \multicolumn{3}{c|}{\textbf{PR}}                        & \multicolumn{3}{c|}{\textbf{CC}}                       \\ \cline{2-14} 
                                          & \textbf{\begin{tabular}[c]{@{}c@{}}runtime \\ (ms)\end{tabular}} & \textbf{GBlst} & \textbf{Ours} & \textbf{Speedup}         & \textbf{GBlst} & \textbf{Ours} & \textbf{Speedup}       & \textbf{GBlst} & \textbf{Ours} & \textbf{Speedup}       & \textbf{GBlst} & \textbf{Ours} & \textbf{Speedup}      \\ \hline
\multirow{2}{*}{\textbf{delaunay\_n14}}   & algorithm                                                            & 3.73           & 1.09          & \textbf{3$\times$}    & 8.35           & 1.99          & \textbf{4$\times$}  & 1.04           & 0.13          & \textbf{8$\times$}  & 1.89           & 0.61          & \textbf{3$\times$} \\ \cline{2-14} 
                                          & kernel                                                           & 2.43           & 0.35          & \textbf{7$\times$}    & 5.11           & 1.08          & \textbf{5$\times$}  & 0.51           & 0.03          & \textbf{14$\times$} & 1.02           & 0.14          & \textbf{7$\times$} \\ \hline
\multirow{2}{*}{\textbf{se}}              & algorithm                                                            & 2.06           & 0.54          & \textbf{4$\times$}    & 4.75           & 1.31          & \textbf{4$\times$}  & 1.22           & 0.17          & \textbf{7$\times$}  & 0.72           & 0.20          & \textbf{4$\times$} \\ \cline{2-14} 
                                          & kernel                                                           & 1.68           & 0.15          & \textbf{12$\times$}   & 2.75           & 0.80          & \textbf{3$\times$}  & 0.56           & 0.03          & \textbf{20$\times$} & 0.27           & 0.06          & \textbf{5$\times$} \\ \hline
\multirow{2}{*}{\textbf{debr}}            & algorithm                                                            & 5.85           & 2.27          & \textbf{3$\times$}    & 28.88          & 15.64         & \textbf{2$\times$}  & 14.39          & 4.20          & \textbf{3$\times$}  & 10.96          & 3.39          & \textbf{3$\times$} \\ \cline{2-14} 
                                          & kernel                                                           & 2.74           & 1.83          & \textbf{2$\times$}    & 14.64          & 11.42         & \textbf{1$\times$}  & 4.51           & 0.52          & \textbf{9$\times$}  & 1.91           & 1.05          & \textbf{2$\times$} \\ \hline
\multirow{2}{*}{\textbf{ash292}}          & algorithm                                                            & 1.45           & 0.02          & \textbf{61$\times$}   & 2.97           & 0.51          & \textbf{6$\times$}  & 0.98           & 0.12          & \textbf{8$\times$}  & 0.65           & 0.17          & \textbf{4$\times$} \\ \cline{2-14} 
                                          & kernel                                                           & 0.83           & 0.01          & \textbf{152$\times$}  & 2.03           & 0.17          & \textbf{12$\times$} & 0.46           & 0.01          & \textbf{35$\times$} & 0.24           & 0.03          & \textbf{9$\times$} \\ \hline
\multirow{2}{*}{\textbf{netz4504\_dual}}  & algorithm                                                            & 2.33           & 0.02          & \textbf{98$\times$}   & 5.32           & 0.88          & \textbf{6$\times$}  & 0.97           & 0.06          & \textbf{16$\times$} & 0.73           & 0.21          & \textbf{4$\times$} \\ \cline{2-14} 
                                          & kernel                                                           & 1.39           & 0.01          & \textbf{221$\times$}  & 3.30           & 0.28          & \textbf{12$\times$} & 0.49           & 0.01          & \textbf{34$\times$} & 0.28           & 0.03          & \textbf{9$\times$} \\ \hline
\multirow{2}{*}{\textbf{minnesota}}       & algorithm                                                            & 6.22           & 0.08          & \textbf{79$\times$}   & 10.48          & 2.66          & \textbf{4$\times$}  & 0.99           & 0.06          & \textbf{17$\times$} & 0.94           & 0.26          & \textbf{4$\times$} \\ \cline{2-14} 
                                          & kernel                                                           & 4.67           & 0.02          & \textbf{243$\times$}  & 6.36           & 0.83          & \textbf{8$\times$}  & 0.47           & 0.01          & \textbf{33$\times$} & 0.38           & 0.04          & \textbf{9$\times$} \\ \hline
\multirow{2}{*}{\textbf{jagmesh6}}        & algorithm                                                            & 7.35           & 0.02          & \textbf{318$\times$}  & 11.50          & 2.32          & \textbf{5$\times$}  & 0.97           & 0.07          & \textbf{15$\times$} & 0.92           & 0.26          & \textbf{3$\times$} \\ \cline{2-14} 
                                          & kernel                                                           & 5.74           & 0.01          & \textbf{1025$\times$} & 7.38           & 0.92          & \textbf{8$\times$}  & 0.47           & 0.02          & \textbf{30$\times$} & 0.40           & 0.04          & \textbf{10$\times$} \\ \hline
\multirow{2}{*}{\textbf{uk}}              & algorithm                                                            & 10.15          & 0.22          & \textbf{46$\times$}   & 15.59          & 2.97          & \textbf{5$\times$}  & 1.00           & 0.07          & \textbf{15$\times$} & 3.15           & 0.93          & \textbf{3$\times$} \\ \cline{2-14} 
                                          & kernel                                                           & 8.29           & 0.05          & \textbf{165$\times$}  & 9.04           & 1.19          & \textbf{8$\times$}  & 0.48           & 0.02          & \textbf{26$\times$} & 1.73           & 0.29          & \textbf{6$\times$} \\ \hline
\multirow{2}{*}{\textbf{whitaker3\_dual}} & algorithm                                                            & 25.27          & 0.06          & \textbf{433$\times$}  & 32.69          & 11.80         & \textbf{3$\times$}  & 1.07           & 0.12          & \textbf{9$\times$}  & 1.93           & 0.67          & \textbf{3$\times$} \\ \cline{2-14} 
                                          & kernel                                                           & 21.67          & 0.02          & \textbf{1414$\times$} & 19.54          & 6.17          & \textbf{3$\times$}  & 0.49           & 0.02          & \textbf{25$\times$} & 0.85           & 0.11          & \textbf{8$\times$} \\ \hline
\multirow{2}{*}{\textbf{rajat07}}         & algorithm                                                            & 4.08           & 0.03          & \textbf{160$\times$}  & 9.82           & 2.12          & \textbf{5$\times$}  & 1.11           & 0.12          & \textbf{9$\times$}  & 1.13           & 0.36          & \textbf{3$\times$} \\ \cline{2-14} 
                                          & kernel                                                           & 2.20           & 0.01          & \textbf{250$\times$}  & 6.05           & 1.15          & \textbf{5$\times$}  & 0.52           & 0.03          & \textbf{16$\times$} & 0.50           & 0.09          & \textbf{6$\times$} \\ \hline
\multirow{2}{*}{\textbf{3dtube}}          & algorithm                                                            & 28.28          & 1.37          & \textbf{21$\times$}   & 8.48           & 6.15          & \textbf{1$\times$}  & 2.54           & 0.30          & \textbf{8$\times$}  & 1.38           & 0.49          & \textbf{3$\times$} \\ \cline{2-14} 
                                          & kernel                                                           & 25.83          & 0.83          & \textbf{31$\times$}   & 6.66           & 5.54          & \textbf{1$\times$}  & 1.71           & 0.12          & \textbf{15$\times$} & 0.82           & 0.28          & \textbf{3$\times$} \\ \hline
\multirow{2}{*}{\textbf{Erdos02}}         & algorithm                                                            & 0.32           & 0.11          & \textbf{3$\times$}    & 1.15           & 0.13          & \textbf{9$\times$}  & 1.09           & 0.20          & \textbf{5$\times$}  & 0.54           & 0.12          & \textbf{5$\times$} \\ \cline{2-14} 
                                          & kernel                                                           & 0.18           & 0.05          & \textbf{4$\times$}    & 0.65           & 0.06          & \textbf{11$\times$} & 0.53           & 0.15          & \textbf{4$\times$}  & 0.17           & 0.03          & \textbf{6$\times$} \\ \hline
\multirow{2}{*}{\textbf{mycielskian9}}    & algorithm                                                            & 0.23           & 0.06          & \textbf{4$\times$}    & 0.81           & 0.07          & \textbf{12$\times$} & 0.95           & 0.07          & \textbf{14$\times$} & 0.46           & 0.10          & \textbf{5$\times$} \\ \cline{2-14} 
                                          & kernel                                                           & 0.10           & 0.01          & \textbf{7$\times$}    & 0.44           & 0.03          & \textbf{17$\times$} & 0.46           & 0.02          & \textbf{23$\times$} & 0.14           & 0.02          & \textbf{7$\times$} \\ \hline
\multirow{2}{*}{\textbf{EX3}}             & algorithm                                                            & 0.13           & 0.13          & \textbf{4$\times$}    & 1.43           & 0.25          & \textbf{6$\times$}  & 0.94           & 0.07          & \textbf{12$\times$} & 0.50           & 0.09          & \textbf{5$\times$} \\ \cline{2-14} 
                                          & kernel                                                           & 0.34           & 0.05          & \textbf{7$\times$}    & 0.82           & 0.08          & \textbf{11$\times$} & 0.46           & 0.04          & \textbf{13$\times$} & 0.15           & 0.03          & \textbf{6$\times$} \\ \hline
\multirow{2}{*}{\textbf{net25}}           & algorithm                                                            & 0.18           & 0.18          & \textbf{3$\times$}    & 2.21           & 0.26          & \textbf{8$\times$}  & 1.40           & 0.19          & \textbf{7$\times$}  & 0.87           & 0.23          & \textbf{4$\times$} \\ \cline{2-14} 
                                          & kernel                                                           & 0.38           & 0.10          & \textbf{4$\times$}    & 1.44           & 0.15          & \textbf{10$\times$}  & 0.83           & 0.13          & \textbf{6$\times$}  & 0.44           & 0.08          & \textbf{5$\times$} \\ \hline
\multirow{2}{*}{\textbf{mycielskian10}}   & algorithm                                                            & 0.09           & 0.06          & \textbf{4$\times$}    & 0.85           & 0.08          & \textbf{10$\times$} & 0.99           & 0.08          & \textbf{12$\times$} & 0.47           & 0.12          & \textbf{4$\times$} \\ \cline{2-14} 
                                          & kernel                                                           & 0.10           & 0.02          & \textbf{6$\times$}    & 0.47           & 0.04          & \textbf{11$\times$} & 0.49           & 0.04          & \textbf{13$\times$} & 0.15           & 0.03          & \textbf{5$\times$} \\ \hline
\end{tabular}
}
\end{table*}

\begin{table*}[]
\caption{SpMV-based graph algorithm performance on Titan V (Volta) GPU.}
\label{tab-spmv-algo-volta}
\centering
\resizebox{1.6\columnwidth}{!}{
\begin{tabular}{|c|c|c|c|c|c|c|c|c|c|c|c|c|c|}
\hline
\multirow{2}{*}{\textbf{Matrix}}            & \textbf{Algorithm}                                                   & \multicolumn{3}{c|}{\textbf{BFS}}                         & \multicolumn{3}{c|}{\textbf{SSSP}}                      & \multicolumn{3}{c|}{\textbf{PR}}                        & \multicolumn{3}{c|}{\textbf{CC}}                        \\ \cline{2-14} 
                                          & \textbf{\begin{tabular}[c]{@{}c@{}}runtime \\ (ms)\end{tabular}} & \textbf{GBlst} & \textbf{Ours} & \textbf{Speedup}         & \textbf{GBlst} & \textbf{Ours} & \textbf{Speedup}       & \textbf{GBlst} & \textbf{Ours} & \textbf{Speedup}       & \textbf{GBlst} & \textbf{Ours} & \textbf{Speedup}       \\ \hline
\multirow{2}{*}{\textbf{delaunay\_n14}}   & algorithm                                                            & 5.17           & 1.75          & \textbf{3$\times$}    & 10.86          & 1.99          & \textbf{5$\times$}  & 1.43           & 0.17          & \textbf{8$\times$}  & 3.09           & 0.67          & \textbf{5$\times$}  \\ \cline{2-14} 
                                          & kernel                                                           & 3.29           & 0.41          & \textbf{8$\times$}    & 6.38           & 0.85          & \textbf{7$\times$}  & 0.65           & 0.04          & \textbf{15$\times$} & 1.67           & 0.14          & \textbf{12$\times$} \\ \hline
\multirow{2}{*}{\textbf{se}}              & algorithm                                                            & 2.59           & 0.63          & \textbf{4$\times$}    & 5.80           & 1.24          & \textbf{5$\times$}  & 1.45           & 0.22          & \textbf{7$\times$}  & 0.97           & 0.24          & \textbf{4$\times$}  \\ \cline{2-14} 
                                          & kernel                                                           & 2.06           & 0.17          & \textbf{12$\times$}   & 3.38           & 0.64          & \textbf{5$\times$}  & 1.45           & 0.22          & \textbf{7$\times$}  & 0.35           & 0.04          & \textbf{9$\times$}  \\ \hline
\multirow{2}{*}{\textbf{debr}}            & algorithm                                                            & 5.32           & 1.26          & \textbf{4$\times$}    & 20.17          & 7.86          & \textbf{3$\times$}  & 8.36           & 5.36          & \textbf{2$\times$}  & 7.37           & 1.63          & \textbf{5$\times$}  \\ \cline{2-14} 
                                          & kernel                                                           & 2.46           & 0.78          & \textbf{3$\times$}    & 9.59           & 5.33          & \textbf{2$\times$}  & 2.21           & 0.19          & \textbf{12$\times$} & 0.98           & 0.41          & \textbf{2$\times$}  \\ \hline
\multirow{2}{*}{\textbf{ash292}}          & algorithm                                                            & 1.91           & 0.03          & \textbf{64$\times$}   & 4.08           & 0.81          & \textbf{5$\times$}  & 1.32           & 0.07          & \textbf{19$\times$} & 0.84           & 0.23          & \textbf{4$\times$}  \\ \cline{2-14} 
                                          & kernel                                                           & 1.09           & 0.01          & \textbf{152$\times$}  & 2.36           & 0.22          & \textbf{11$\times$} & 0.62           & 0.01          & \textbf{43$\times$} & 0.32           & 0.03          & \textbf{10$\times$} \\ \hline
\multirow{2}{*}{\textbf{netz4504\_dual}}  & algorithm                                                            & 3.22           & 0.03          & \textbf{115$\times$}  & 7.38           & 1.37          & \textbf{5$\times$}  & 1.40           & 0.08          & \textbf{19$\times$} & 0.87           & 0.27          & \textbf{3$\times$}  \\ \cline{2-14} 
                                          & kernel                                                           & 1.89           & 0.01          & \textbf{264$\times$}  & 4.23           & 0.36          & \textbf{12$\times$} & 0.57           & 0.02          & \textbf{33$\times$} & 0.36           & 0.04          & \textbf{9$\times$}  \\ \hline
\multirow{2}{*}{\textbf{minnesota}}       & algorithm                                                            & 8.60           & 0.09          & \textbf{96$\times$}   & 14.48          & 3.15          & \textbf{5$\times$}  & 1.33           & 0.09          & \textbf{15$\times$} & 1.20           & 0.36          & \textbf{3$\times$}  \\ \cline{2-14} 
                                          & kernel                                                           & 6.34           & 0.02          & \textbf{282$\times$}  & 8.36           & 0.83          & \textbf{10$\times$} & 0.62           & 0.02          & \textbf{32$\times$} & 0.48           & 0.05          & \textbf{9$\times$}  \\ \hline
\multirow{2}{*}{\textbf{jagmesh6}}        & algorithm                                                            & 9.82           & 0.03          & \textbf{349$\times$}  & 8.36           & 0.83          & \textbf{10$\times$} & 1.33           & 0.07          & \textbf{20$\times$} & 1.09           & 0.35          & \textbf{3$\times$}  \\ \cline{2-14} 
                                          & kernel                                                           & 6.69           & 0.01          & \textbf{824$\times$}  & 9.07           & 0.95          & \textbf{10$\times$}  & 0.58           & 0.02          & \textbf{33$\times$} & 0.51           & 0.05          & \textbf{10$\times$}  \\ \hline
\multirow{2}{*}{\textbf{uk}}              & algorithm                                                            & 13.39          & 0.25          & \textbf{53$\times$}   & 20.57          & 4.61          & \textbf{4$\times$}  & 1.31           & 0.09          & \textbf{15$\times$} & 0.51           & 0.05          & \textbf{10$\times$}  \\ \cline{2-14} 
                                          & kernel                                                           & 11.85          & 0.07          & \textbf{175$\times$}  & 11.43          & 1.29          & \textbf{9$\times$}  & 0.63           & 0.02          & \textbf{31$\times$} & 3.26           & 0.26          & \textbf{13$\times$} \\ \hline
\multirow{2}{*}{\textbf{whitaker3\_dual}} & algorithm                                                            & 30.97          & 0.08          & \textbf{414$\times$}  & 40.67          & 10.11         & \textbf{4$\times$}  & 1.33           & 0.16          & \textbf{8$\times$}  & 2.59           & 0.64          & \textbf{4$\times$}  \\ \cline{2-14} 
                                          & kernel                                                           & 26.14          & 0.02          & \textbf{1344$\times$} & 23.82          & 4.46          & \textbf{5$\times$}  & 0.64           & 0.02          & \textbf{33$\times$} & 1.08           & 0.12          & \textbf{10$\times$}  \\ \hline
\multirow{2}{*}{\textbf{rajat07}}         & algorithm                                                            & 5.20           & 0.03          & \textbf{165$\times$}  & 11.96          & 2.37          & \textbf{5$\times$}  & 1.34           & 0.16          & \textbf{8$\times$}  & 1.46           & 0.41          & \textbf{4$\times$}  \\ \cline{2-14} 
                                          & kernel                                                           & 3.13           & 0.01          & \textbf{339$\times$}  & 6.96           & 1.01          & \textbf{7$\times$}  & 0.62           & 0.04          & \textbf{18$\times$} & 0.61           & 0.08          & \textbf{8$\times$}  \\ \hline
\multirow{2}{*}{\textbf{3dtube}}          & algorithm                                                            & 17.65          & 1.01          & \textbf{18$\times$}   & 7.52           & 5.72          & \textbf{1$\times$}  & 2.08           & 0.34          & \textbf{6$\times$}  & 1.21           & 0.38          & \textbf{3$\times$}  \\ \cline{2-14} 
                                          & kernel                                                           & 15.13          & 0.39          & \textbf{39$\times$}   & 5.06           & 4.89          & \textbf{1$\times$}  & 1.14           & 0.08          & \textbf{15$\times$} & 0.61           & 0.13          & \textbf{5$\times$}  \\ \hline
\multirow{2}{*}{\textbf{Erdos02}}         & algorithm                                                            & 0.43           & 0.13          & \textbf{3$\times$}    & 1.53           & 0.17          & \textbf{9$\times$}  & 1.43           & 0.26          & \textbf{6$\times$}  & 0.62           & 0.16          & \textbf{4$\times$}  \\ \cline{2-14} 
                                          & kernel                                                           & 0.26           & 0.06          & \textbf{4$\times$}    & 0.84           & 0.07          & \textbf{11$\times$} & 0.69           & 0.18          & \textbf{4$\times$}  & 0.22           & 0.04          & \textbf{6$\times$}  \\ \hline
\multirow{2}{*}{\textbf{mycielskian9}}    & algorithm                                                            & 0.29           & 0.07          & \textbf{4$\times$}    & 1.08           & 0.11          & \textbf{10$\times$} & 1.29           & 0.08          & \textbf{17$\times$} & 0.59           & 0.14          & \textbf{4$\times$}  \\ \cline{2-14} 
                                          & kernel                                                           & 0.14           & 0.02          & \textbf{9$\times$}    & 0.58           & 0.04          & \textbf{16$\times$} & 0.60           & 0.03          & \textbf{25$\times$} & 0.19           & 0.03          & \textbf{7$\times$}  \\ \hline
\multirow{2}{*}{\textbf{EX3}}             & algorithm                                                            & 0.72           & 0.21          & \textbf{3$\times$}    & 1.68           & 0.24          & \textbf{7$\times$}  & 1.09           & 0.10          & \textbf{10$\times$} & 0.56           & 0.13          & \textbf{4$\times$}  \\ \cline{2-14} 
                                          & kernel                                                           & 0.46           & 0.07          & \textbf{7$\times$}    & 0.97           & 0.09          & \textbf{11$\times$} & 0.57           & 0.05          & \textbf{12$\times$} & 0.21           & 0.03          & \textbf{8$\times$}  \\ \hline
\multirow{2}{*}{\textbf{net25}}           & algorithm                                                            & 0.69           & 0.23          & \textbf{3$\times$}    & 2.27           & 0.29          & \textbf{8$\times$}  & 1.36           & 0.25          & \textbf{5$\times$}  & 0.93           & 0.32          & \textbf{3$\times$}  \\ \cline{2-14} 
                                          & kernel                                                           & 0.46           & 0.13          & \textbf{4$\times$}    & 1.29           & 0.13          & \textbf{10$\times$}  & 0.67           & 0.16          & \textbf{4$\times$}  & 0.39           & 0.11          & \textbf{4$\times$}  \\ \hline
\multirow{2}{*}{\textbf{mycielskian10}}   & algorithm                                                            & 0.31           & 0.07          & \textbf{4$\times$}    & 1.13           & 0.14          & \textbf{8$\times$}  & 1.31           & 0.09          & \textbf{14$\times$} & 0.56           & 0.14          & \textbf{4$\times$}  \\ \cline{2-14} 
                                          & kernel                                                           & 0.14           & 0.02          & \textbf{6$\times$}    & 0.56           & 0.05          & \textbf{10$\times$} & 0.64           & 0.04          & \textbf{15$\times$} & 0.19           & 0.04          & \textbf{4$\times$}  \\ \hline
\end{tabular}
}
\end{table*}

In \textbf{\texttt{bmv\_bin\_bin\_full()}}, the vector input is binarized with the same tile dimension of the binary adjacency matrix. Compared to \textbf{\texttt{bmv\_bin\_full\_full()}}, it requires less vector load bandwidth per matrix-vector multiplication. The inner product of each bit-row is done by bit-wise AND and population count of the resulting bit-row. In the 521 binary matrices, the runtime speedup of \textbf{\texttt{bmv\_bin\_bin\_full()}} over cuSPARSE's full-precision CSR SpMV is shown in Figure~\ref{bmv-bin-bin-full-pascal} and~\ref{bmv-bin-bin-full-volta}. On GTX1080, \textbf{\texttt{bmv\_bin\_bin\_full()}} achieves an average speedup of 2.06$\times$ in B2SR-4, 2.36$\times$ in B2SR-8, 2.22$\times$ in B2SR-16, and 2.97$\times$ in B2SR-32. The max speedup over baseline is 38$\times$ that appears at matrix \textit{ins2} with B2SR-32. On Titan V, \textbf{\texttt{bmv\_bin\_bin\_full()}} achieves an average speedup of 1.72$\times$ in B2SR-4, 1.84$\times$ in B2SR-8, 1.92$\times$ in B2SR-16, and 2.32$\times$ in B2SR-32. The max speedup over baseline is 25$\times$ that appears at matrix \textit{vsp\_c-30\_data\_data} with B2SR-32. 

For \textbf{\texttt{bmv\_bin\_full\_full()}}, the performance is present in Figure~\ref{bmv-bin-full-full-pascal} and~\ref{bmv-bin-full-full-volta}. Unlike \textbf{\texttt{bmv\_bin\_bin\_full()}}, the average performance gain decreases when enlarged the tile size. On GTX1080, it achieves an average speedup of 2.06$\times$ in B2SR-4, 1.92$\times$ in B2SR-8, 1.43$\times$ in B2SR-16, and 0.92$\times$ in B2SR-32. The most significant speedup happens at \textit{vsp\_south31\_slptsk} with B2SR-4, yielding 18$\times$. On Titan V, it achieves an average speedup of 1.88$\times$ in B2SR-4, 1.71$\times$ in B2SR-8, 1.27$\times$ in B2SR-16, and 0.81$\times$ in B2SR-32. The most significant speedup happens at matrix \textit{ins2} with B2SR-4, with 14$\times$ acceleration.

\textbf{BMM} In BMM, we implement \textbf{\texttt{bmm\_bin\_bin\_sum()}} to support the SpGEMM kernels (see Table~\ref{bmm-scheme}). Figure~\ref{bmm-bin-bin-sum-pascal} and~\ref{bmm-bin-bin-sum-volta} presents the performance of the BMM kernel compared to cuSPARSE's \textbf{\texttt{cusparseScsrgemm()}}. On GTX1080, it achieves an average speedup of 33.96$\times$ in B2SR-4, 27.84$\times$ in B2SR-8, 17.81$\times$ in B2SR-16, and 10.22$\times$ in B2SR-32. The maximum speedup is 6555$\times$ that happens at matrix \textit{ins2} with B2SR-4. On Titan V, the performance gain is moderate compared to GTX1080. We accounts the reason for cuSPARSE's APIs have better performance gain on Volta than Pascal, while our implementation perform similar or evenly slightly poor on Volta than on Pascal. \textbf{\texttt{bmm\_bin\_bin\_sum()}} on Titan V achieves an average speedup of 5.34$\times$ in B2SR-4, 3.65$\times$ in B2SR-8, 9.03$\times$ in B2SR-16, and 12.25$\times$ in B2SR-32. Interestingly, the most significant speedup is 3243$\times$ also happens at matrix \textit{ins2} but with B2SR-32 instead of B2SR-4 as on GTX1080.

\begin{table}[]
\caption{SpGEMM-based graph algorithm performance on GTX1080 (Pascal) and Titan V (Volta) GPU.}
\label{tc-pascal-volta}
\centering
\resizebox{\columnwidth}{!}{
\begin{tabular}{|c|c|c|c|c|c|c|}
\hline
\multirow{2}{*}{\textbf{Matrix}}   & \multicolumn{3}{c|}{\textbf{TC runtime (ms) on Pascal}} & \multicolumn{3}{c|}{\textbf{TC runtime (ms) on Volta}}   \\ \cline{2-7} 
                                   & \textbf{GBlst} & \textbf{Ours} & \textbf{Speedup}       & \textbf{GBlst} & \textbf{Ours} & \textbf{Speedup}        \\ \hline
\textbf{delaunay\_n14}             & 0.14           & 0.06          & \textbf{2$\times$}  & 0.11           & 0.13          & \textbf{1$\times$}   \\ \hline
\textbf{se}                        & 0.14           & 0.03          & \textbf{4$\times$}  & 0.11           & 0.02          & \textbf{5$\times$}   \\ \hline
\textbf{debr}                      & 3.51           & 0.26          & \textbf{13$\times$} & 1.07           & 0.09          & \textbf{12$\times$}  \\ \hline
\textbf{sstmodel}                  & 0.51           & 0.03          & \textbf{20$\times$} & 0.66           & 0.03          & \textbf{21$\times$}  \\ \hline
\textbf{jagmesh2}                  & 0.29           & 0.01          & \textbf{22$\times$} & 0.10           & 0.02          & \textbf{4$\times$}   \\ \hline
\textbf{lock2232}                  & 0.51           & 0.03          & \textbf{19$\times$} & 0.62           & 0.03          & \textbf{22$\times$}  \\ \hline
\textbf{ramage02}                  & 12.59          & 0.53          & \textbf{24$\times$} & 3.96           & 0.37          & \textbf{11$\times$}  \\ \hline
\textbf{s4dkt3m2}                  & 3.83           & 0.15          & \textbf{26$\times$} & 1.05           & 0.06          & \textbf{18$\times$}  \\ \hline
\textbf{opt1}                      & 7.03           & 0.27          & \textbf{26$\times$} & 2.30           & 0.21          & \textbf{11$\times$}  \\ \hline
\textbf{trdheim}                   & 3.84           & 0.08          & \textbf{49$\times$} & 1.29           & 0.06          & \textbf{23$\times$}  \\ \hline
\textbf{3dtube}                    & 151.89         & 2.91          & \textbf{52$\times$} & 79.49          & 2.95          & \textbf{27$\times$}  \\ \hline
\textbf{mycielskian12}             & 22.47          & 4.63          & \textbf{5$\times$}  & 12.51          & 4.47          & \textbf{3$\times$}   \\ \hline
\textbf{Erdos02}                   & 7.37           & 0.31          & \textbf{23$\times$} & 3.62           & 0.57          & \textbf{6$\times$}   \\ \hline
\textbf{mycielskian9}              & 0.48           & 0.08          & \textbf{6$\times$}  & 0.35           & 0.08          & \textbf{4$\times$}   \\ \hline
\textbf{mycielskian13}             & 93.21          & 20.37         & \textbf{5$\times$}  & 48.87          & 18.87         & \textbf{3$\times$}   \\ \hline
\textbf{vsp\_c-60\_data\_cti\_cs4} & 139.41         & 9.43          & \textbf{15$\times$} & 72.32          & 5.19          & \textbf{14$\times$} \\ \hline
\end{tabular}
}
\end{table}

\subsection{Graph Algorithms}
We compare the implemented B2SR-based Bit-GraphBLAS algorithms with GraphBLAST~\cite{yang2020graphblast}, a state-of-the-art GPU-based GraphBLAS framework. GraphBLAST switches between sparse and dense vector/matrix computation depending on sparsity degree across algorithm interactions with optimized CUDA kernels. For the GraphBLAST configuration, we use the same environment (e.g., CUDA Runtime 9.1) as indicated on the Github repository~\cite{graphblastgithub}. The graph algorithms are implemented following the GraphBLAS convention. For iteration-based algorithms, such as BFS, SSSP, PR, and CC, each iteration contains a frontier vector that performs a matrix-vector multiplication with the adjacency matrix and several element-wise scalar operations. They are used to update the frontier vector for indicating neighbor aggregation in each iteration, through the mathematical semi-ring operation. The number of iterations depends on when the algorithm is converged at runtime. 

Since matrix-vector multiplication is the major performance concern per iteration ($>$80\% of the workload), in Bit-GraphBLAS, our major goal is to improve the efficiency of SpMV through our B2SR based BMV kernel. Table~\ref{tab-spmv-algo-pascal} and~\ref{tab-spmv-algo-volta} list the algorithm and kernel execution latency in ms for the proposed Bit-GraphBLAS with respect to GraphBLAST. As can be seen, under both conditions, Bit-GraphBLAS achieves considerable speedups through B2SR and the strong bit computation capability of modern GPUs.
 

For Bit-GraphBLAS, in the 521 binary matrices dataset, the patterns with better performance fall into three main categories: \textbf{diagonal}, \textbf{block}, and \textbf{stripe} (reference the classification in Table~\ref{mat-pattern}). The performance of matrices from the three categories are shown in Table~\ref{tab-spmv-algo-pascal} and~\ref{tab-spmv-algo-volta}. In the subset of matrices, \textit{delaunay\_n14}, \textit{se}, and \textit{debr} are \textbf{stripe} patterns; \textit{Erdos02}, \textit{mycielskian9}, \textit{EX3}, \textit{net25}, and \textit{mycielskian10} are \textbf{block} patterns; \textit{ash292}, \textit{netz4504\_dual}, \textit{minnesota}, \textit{jagmesh6}, \textit{uk}, \textit{whitaker3\_dual}, \textit{rajat07}, \textit{3dtube} are \textbf{diagonal} patterns. In the algorithm evaluation, \textbf{BFS} relies on the kernel \textbf{\texttt{bmv\_bin\_bin\_bin\_masked()}} with boolean semiring. On Pascal, \textbf{diagonal} pattern matrices can achieve up to 433$\times$ acceleration in the whole algorithm and 1414$\times$ in kernel; On Volta, it achieves up to 349$\times$ speedup to GraphBLAST in algorithm and 1344$\times$ in kernel. Matrices in \textbf{stripe} and \textbf{block} generally perform a moderate speedup from 2$\times$ to 7$\times$ on both GPU architectures. Other three SpMV-based algorithms (\textbf{SSSP}, \textbf{PR}, and \textbf{CC}) are fulfilled by \textbf{\texttt{bmv\_bin\_full\_full()}} with relaxation and extension (details are described in Section~\ref{graph-algo-section}). They mainly achieve a acceleration over GraphBLAST with no more than 20$\times$ algorithm-wise and 40$\times$ kernel-wise. 

In Table~\ref{tc-pascal-volta}, we demonstrate the performance improvement of TC algorithm, which is essentially a one-time execution of the \textbf{\texttt{bmm\_bin\_sum\_masked()}} kernel. On both Pascal and Volta, we still have \textbf{diagonal} patterns with the best performance. It can achieve up to 52$\times$ on Pascal and 27$\times$ on Volta GPUs. 

It is noteworthy that the same matrix can often find lower GraphBLAST runtime on Volta than on Pascal. A similar effect can be found in both kernel and algorithm evaluations. For example, the \textit{3dtube} runs 151.89 (ms) on Pascal but only 79.49 (ms) on Volta. Nevertheless, Bit-GraphBLAS can have a larger or similar runtime on Volta than on Pascal GPUs. We have a 0.04 (ms) increase in runtime for the \textit{3dtube} case. We attribute the effect to that Volta has updated the warp execution model and eliminated implicit warp synchronous. This poses a little performance slowdown of binary intrinsics like \textbf{\texttt{\_\_shfl\_sync()}} and \textbf{\texttt{\_\_ballot\_sync()}} compared to the non-synchronizing \textbf{\texttt{\_\_shfl()}} and \textbf{\texttt{\_\_ballot()}} in Pascal GPUs. 

\section{Discussion}
\textbf{Limitations of Bit-GraphBLAS:} As Bit-GraphBLAS relies on the bit-operation and bit-data, it only directly applies to homogeneous graphs (i.e., the adjacency matrix is a binary matrix). Throughout the SuiteSparse data collection, we have observed that $\sim$20\% of graphs are homogeneous that can directly benefit from Bit-GraphBLAS. These graphs cover a wide range of domains including mathematics, power-grid, physics, electronics, material science, economics, thermal, fluid dynamics, etc. Additionally, as the weights for many heterogeneous graphs can be expressed by integers or fixed-points (e.g., through normalization), similar to the recent effort decomposing a quantized-neural-network into several concurrent binary-neural-networks for acceleration~\cite{feng2021apnn}, Bit-GraphBLAS can also be extended to support heterogeneous graphs with short bit-width. We set this as future work.

\textbf{Platform Portability:} Although in the evaluation we showcase Bit-GraphBLAS on NVIDIA GPUs due to hardware availability, the bit intrinsics that Bit-GraphBLAS relies on, such as \_\_popc(), \_\_shfl(), \_\_ballot(), and \_\_brev() are also available (despite using different names) in other GPUs and CPUs (like AMD's GPU and X86 CPUs). Therefore, we did not see significant challenges in supporting Bit-GraphBLAS on alternative hardware platforms (e.g., AMD GPUs through HIPIFY~\cite{hipify}).

\section{Conclusion}
We present Bit-GraphBLAS, a linear algebra-based graph framework that utilizes a Bit-Block Compressed Sparse Row (B2SR) format and bit manipulation primitives on GPUs to enable dense bit-operations on bit tiles within large sparse adjacency matrices. We explore different tile size configurations from 4$\times $4 to 32$\times$32, and suitable bit-packing types. We implement BMV and BMM schemes to support parse kernel operations and demonstrate their effectiveness on graph algorithms by reducing memory footprint. The result shows significant performance gain over full-CSR-based GPU graph frameworks. In sum, the novel storage format and algorithms compress the graphs' storage and accelerate the linear algebra kernels SpMV and SpGEMM through finer-grained bit-wise parallelism. 

\section*{Acknowledgment}
We thank the anonymous reviewers for their helpful comments. This work was partially supported by the U.S. Department of Energy, Office of Science, Office of Advanced Scientific Computing Research, ComPort: Rigorous Testing Methods to Safeguard Software Porting, under Award Number 78284. The evaluation platforms were supported by the U.S. DOE Office of Science, Office of Advanced Scientific Computing Research, under award 66150: "CENATE - Center for Advanced Architecture Evaluation". The work was additionally supported by NSF under Grants CNS-1717425, CCF-1703487, CCF-2028850. Any opinions, findings, conclusions, or recommendations expressed in this material are those of the authors and do not necessarily reflect
the views of NSF or DOE.
\\

\bibliographystyle{IEEEtran}
\bibliography{refs}

\end{document}